*I*nternational

*V*irtual

*O*bservatory

*A*lliance

# IVOA Architecture

# Version 1.0

## *IVOA Note 2010-11-23*

**Editors**

Christophe Arviset

**Author(s):**

Christophe Arviset, Severin Gaudet

**and the IVOA Technical Coordination Group** (tcg@ivoa.net):

## Abstract

This note describes the technical architecture of the IVOA. The description is decomposed into three levels. Level 0 is a general, high level summary of the IVOA Architecture. Level 1 provides more details about components and functionalities, still without being overly technical. Finally, Level 2 displays how the IVOA standards fit into the IVOA Architecture.

## Status of This Document

*This IVOA Note reflects extensive discussion within the IVOA Technical Coordination Group, and it has been formally endorsed by the IVOA Executive Committee. This document accurately reflects the IVOA technical architecture, and will be updated as this architecture and the associated standards evolve.*

## Contents

# 1 IVOA Architecture Level 0

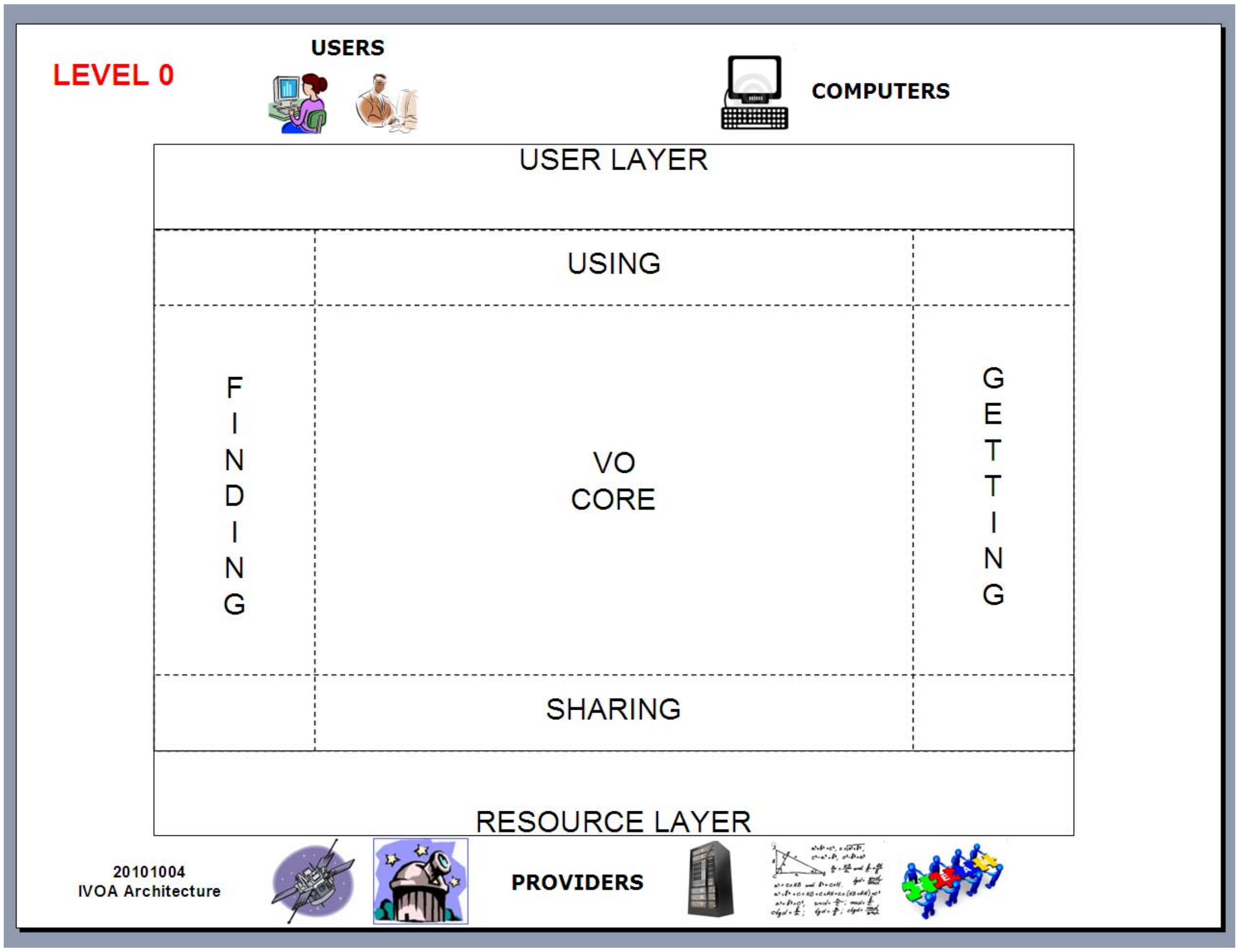

**Figure 1 : IVOA Architecture Level 0**

Astronomy produces large amounts of data of many kinds, coming from various sources: science space missions, ground based telescopes, theoretical models, compilation of results, etc. These data are usually managed by large data centres or smaller teams. These *providers* provide the scientific community with data and / or computing services through the Internet. This is the *Resource Layer*.

The “consumers” of these data and computing services, be it individual researchers, research teams or computer systems, interact with the *User Layer*.

The *Virtual Observatory* is the necessary “middle layer” framework connecting the Resource Layer to the User Layer in a seamless and transparent manner. Like the web which enables end users and machines to access transparently documents and services wherever and however they are stored, the VO enables the astronomical scientific community to access astronomical resources wherever and however they are stored by the astronomical data and services providers. The VO provides a technical framework for the providers to share their data and services (“Sharing”), and allowing users to find (“Finding”) these resources, to get them (“Getting”) and to use them (“Using”). To enable these functionalities, the definition of some core astronomically-oriented standards (“VO Core”) is also necessary.

# 2 IVOA Architecture Level 1

Level 1 of the IVOA architecture is an extension to the Level 0, displaying more details about the functionalities and building blocks within the different layers. For completeness, part of the description is repeated from the Level 0 one, so the Level 1 description can be used as a self-contained block.

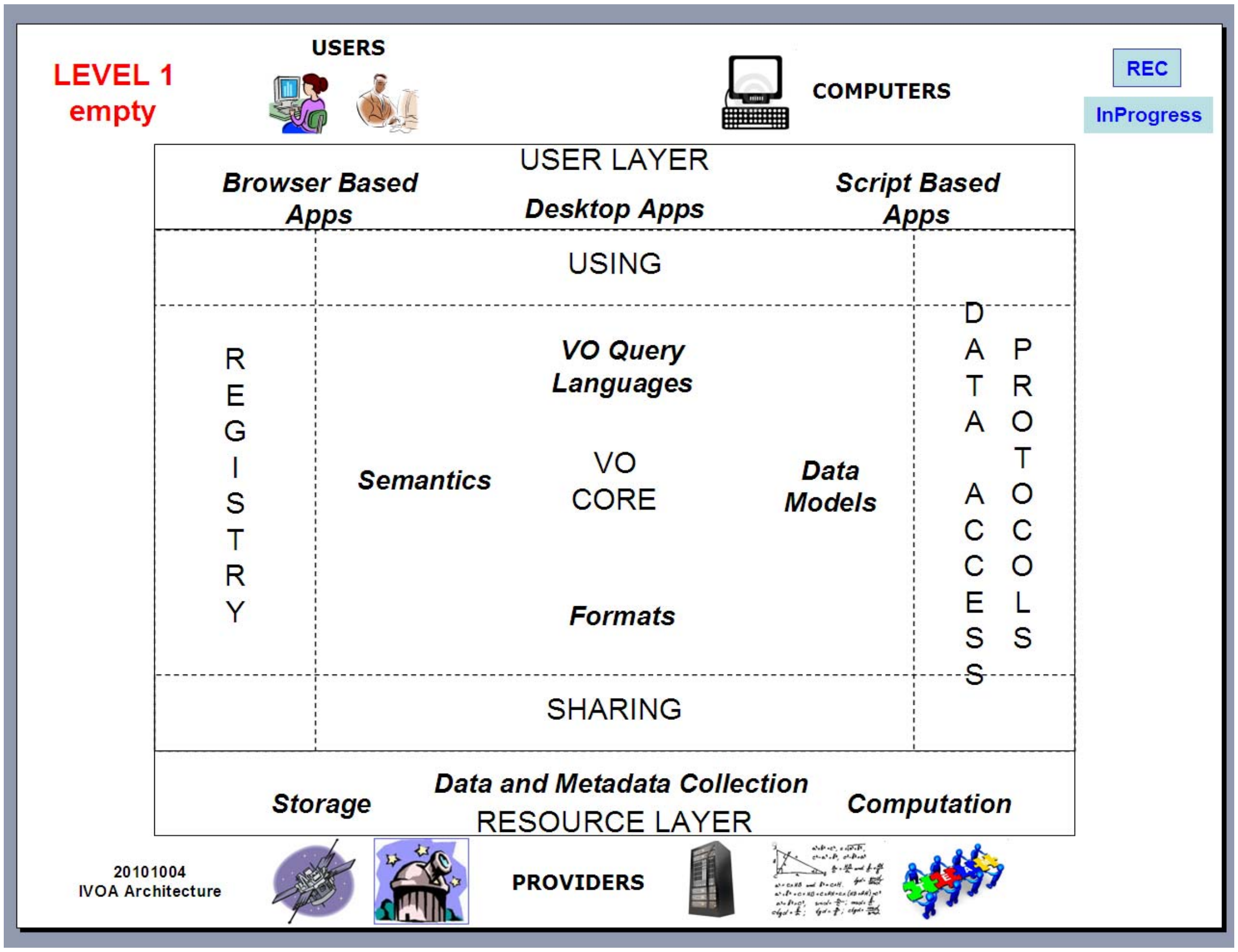

**Figure 2: IVOA Architecture Level 1**

Astronomy produces large amounts of data of many kinds, coming from various sources: science space missions, ground based telescopes, theoretical models, compilation of results, etc. These data are usually managed by large data centres or smaller teams. These *providers* provide the scientific community with data and / or computing services through the Internet. These resources provided can be:

1. data collections (images, spectra, catalogues, time series, theoretical models, etc.) with their associated descriptive metadata and access services.
2. storage services for users and for processing
3. computing services to process data from data collections and from users

This is the *Resource Layer*.

The “consumers” of these data and computing services, be it individual researchers, research teams or computer systems, interact with the *User Layer* of the IVOA architecture.. These interactions can be through browser based applications in a typical web browser, standalone desktop applications or scriptable applications that can be used in automatic and batch modes by a computer.

The *Virtual Observatory* is the necessary “middle layer” framework which connects the Resource Layer to the User Layer in a seamless and transparent manner. Like the web which enables end users and machines to access transparently documents and services wherever and however they are stored, the VO enables the astronomical scientific community to access astronomical resources wherever and however they are stored by the astronomical data and services providers.

The VO provides a technical framework for the *providers* and the *consumers* to share their data and services (“Sharing”).

Registries function as the “yellow pages” of the VO, collecting metadata about data resources and information services into a queryable database. Like the VO resources and services themselves, the registry is also distributed. Replicas exist around the network, both for redundancy and for more specialized collections.

Access to data and metadata collection is done through Data Access Protocols, which specify a uniform way of getting data and metadata from various different providers.

To allow these functionalities, the definition of some core astronomically-oriented standards (“VO Core”) is necessary. In particular, defining common formats and data models and using common semantics is required to have a uniform and common description of astronomical datasets so they can become interoperable and queryable through standard query languages to enable cross analysis amongst various datasets.

Additional standards are required within the User Layer to enable user authentication to proprietary datasets and storage elements as well as interoperability amongst VO applications (“Using”).

# 3 IVOA Architecture Level 2

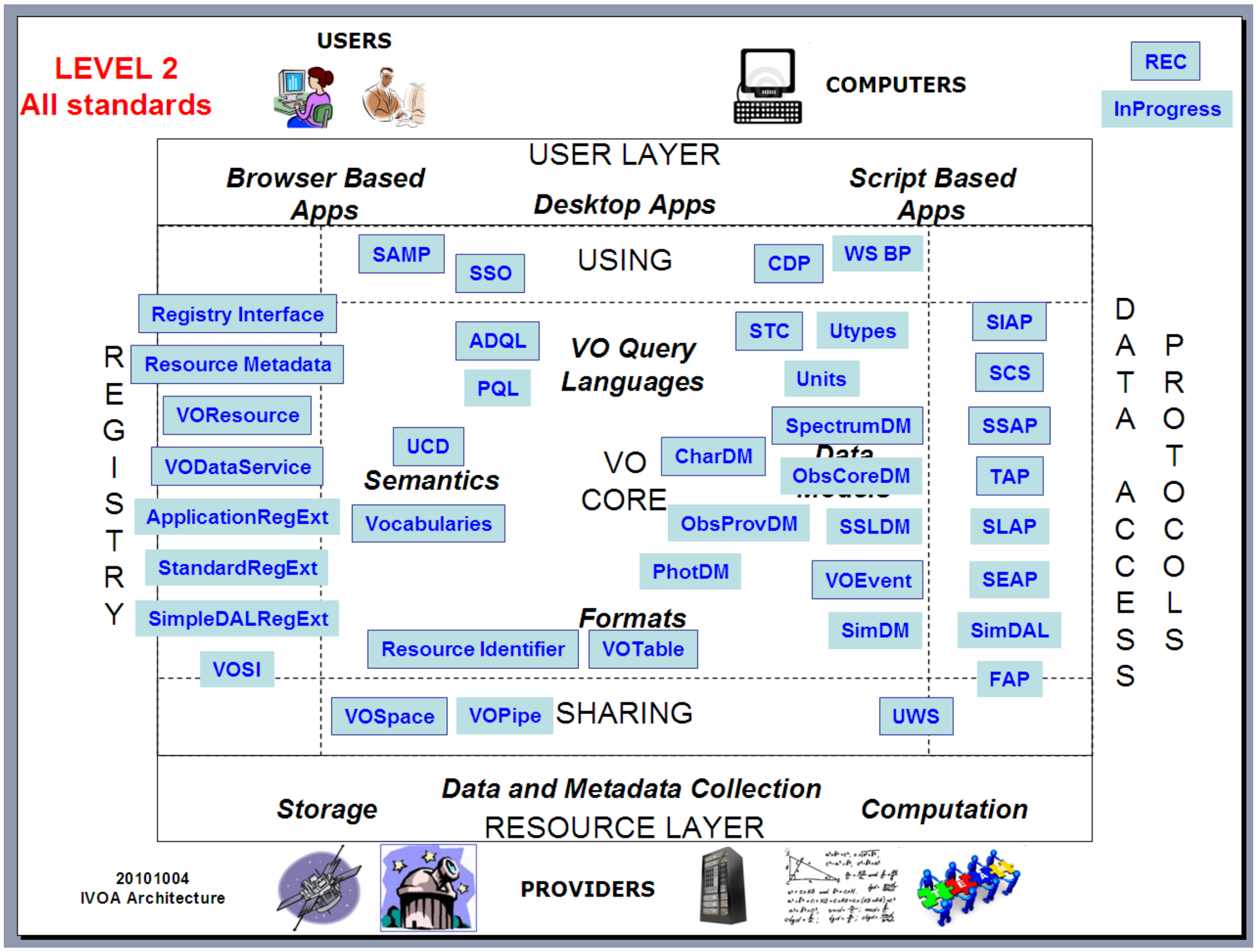

**Figure 3: IVOA Architecture Level 2**

Level 2 of the IVOA Architecture is similar to the Level 1, but adds all the IVOA standards in their corresponding layer. Some standards have already been approved and recommended (blue boxes *with* an outer line) while others are still being worked on (blue boxes *without* an outer line).

Note that this list (and standard status) will naturally evolve with time. More standards will be approved and recommended. Additionally, as driven by science use cases, new standards will be identified and added to that Figure 3.

The following sections provide a summary description of each individual standard, including where it fits into the IVOA architecture and its possible links with other IVOA standards.

# 4 Finding / Registry Standards

The IVOA Registry enables users and applications in the User Layer to discover data and metadata collections, as well services in the Resource Layer. The Registry contains descriptions of VO resources. A *resource* is a general term referring to a VO element that can be described in terms of who curates or maintains it and which can be given a name and a unique Resource Identifier. A resource can be of various types: a data or metadata collection, a computing or storage element, an application, a data and metadata access service, etc.
These resource descriptions are defined by the various Registry standards described in the following sections.

Figure 4 shows how these various Registry standards relate to each other.

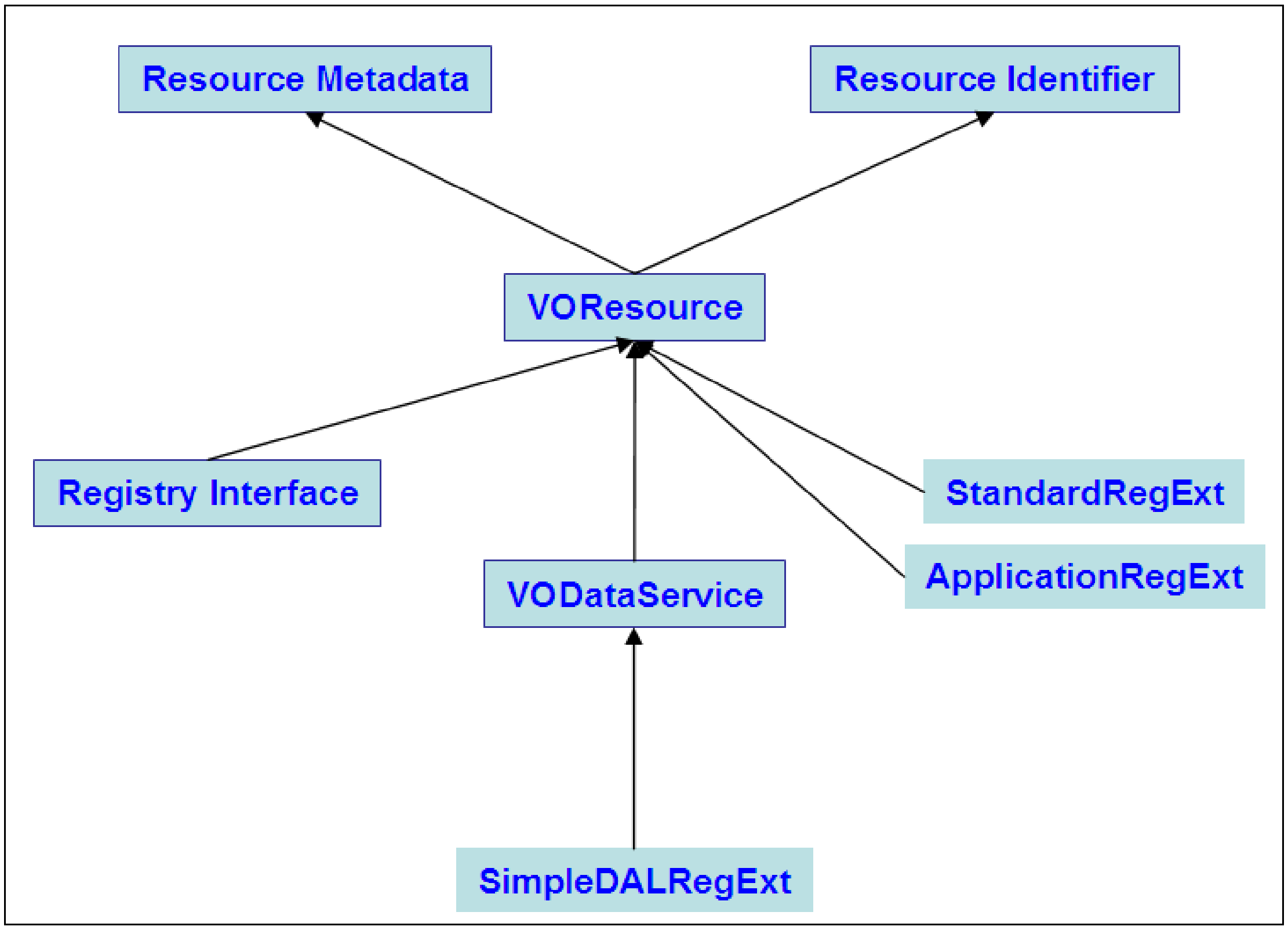

**Figure 4: Relationship between Registry Standards**

## 4.1 Resource Metadata

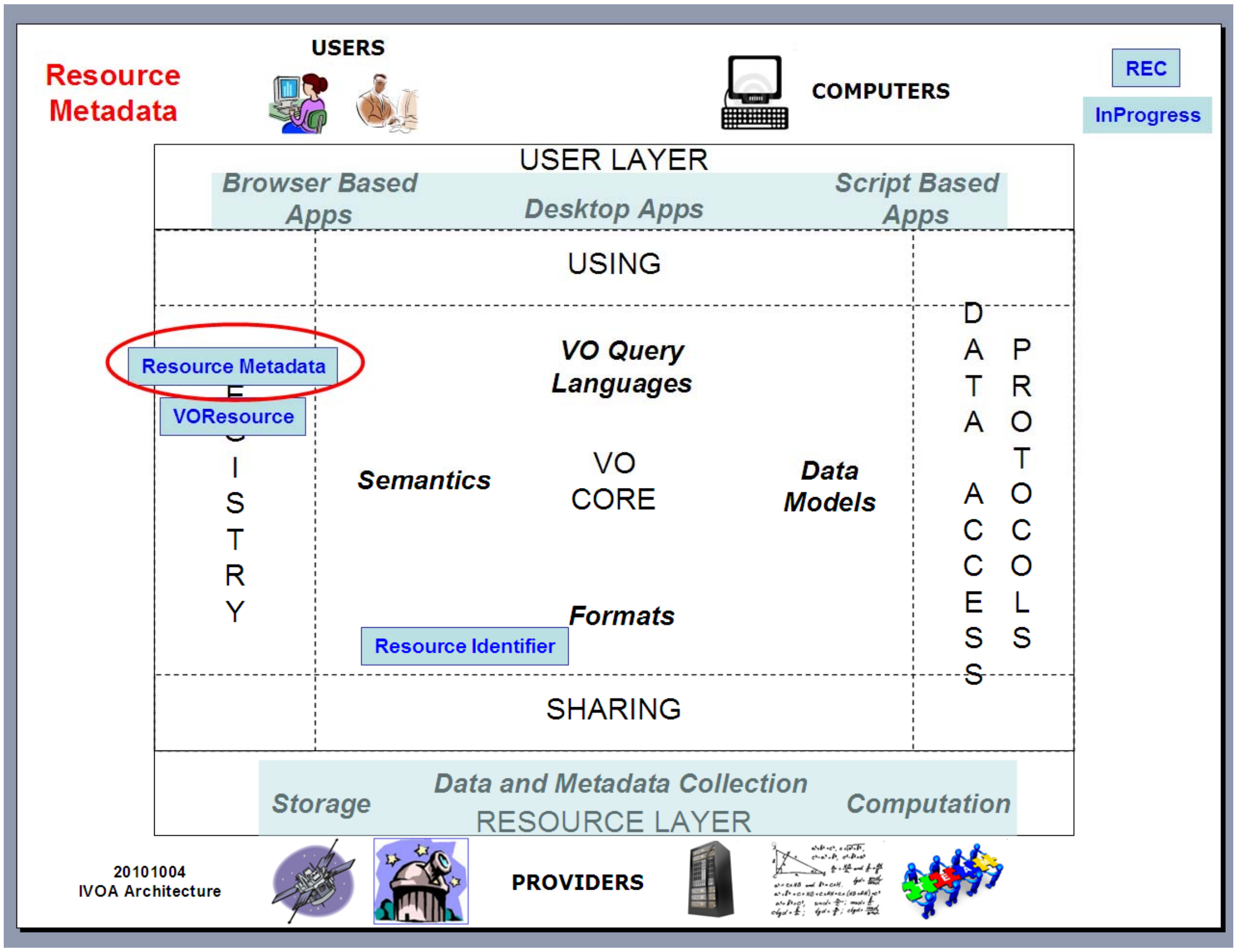

The IVOA Registry enables users and applications in the User Layer to discover data and metadata collections, as well services in the Resource Layer. A *resource* is a general term referring to a VO element that can be described in terms of who curates or maintains it and which can be given a name and a unique Resource Identifier. A resource can be of various types: a data or metadata collection, a computing or storage element, an application, a data and metadata access service, etc.

The Resource Metadata standard represents the essential capability to describe what data and computational facilities are available where, and once identified, how to use them. The data themselves have associated metadata (e.g., FITS keywords), and similarly we require metadata about data collections and data services so that VO users can easily find information of interest.

Resource Metadata became an IVOA Recommended standard in March 2007.

## 4.2 VOResource

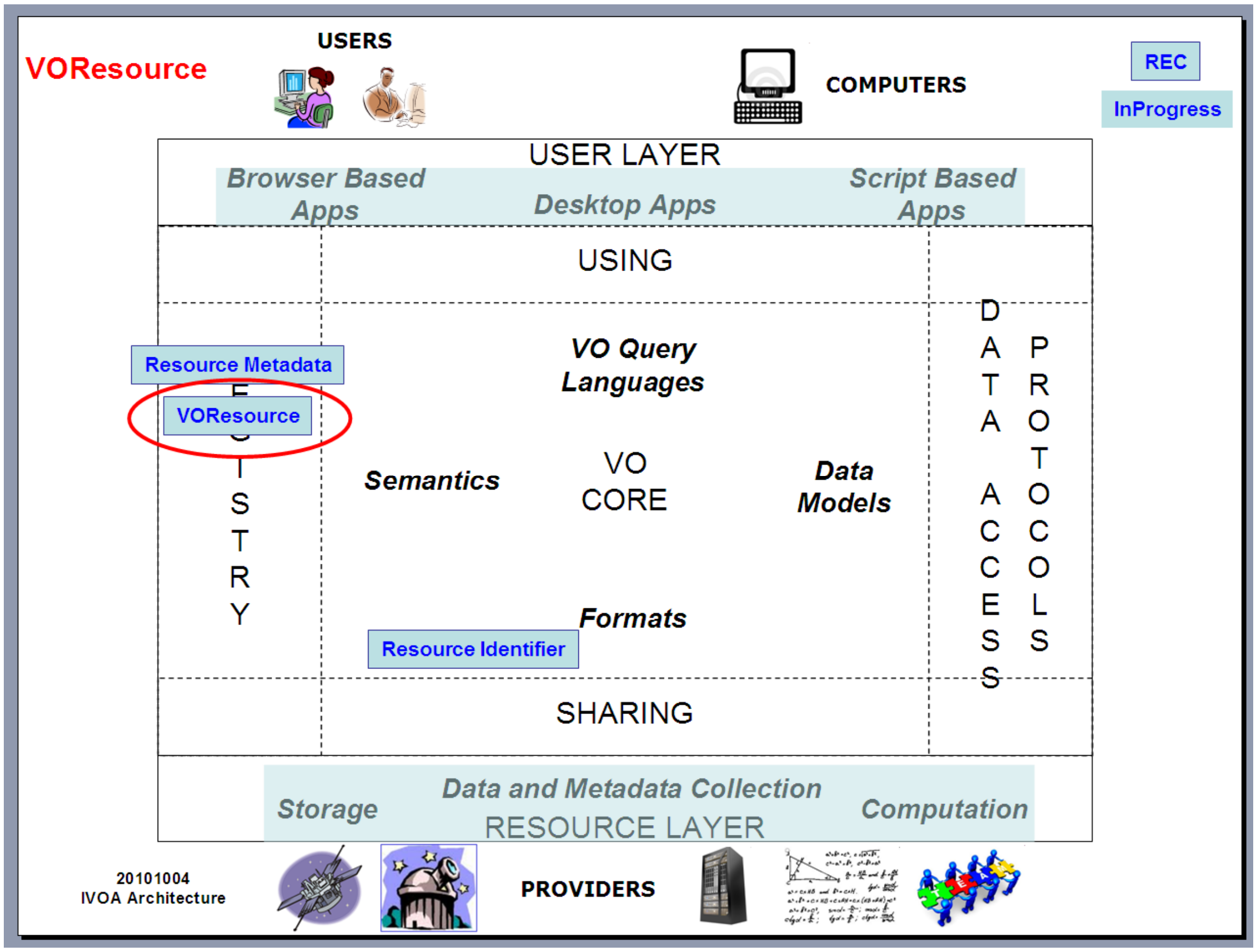

The IVOA Registry enables users and applications in the User Layer to discover data and metadata collection, as well services in the Resource Layer. The Registry contains Resources which are defined by the Resource Metadata standard and get a unique Resource Identifier. VOResource standard is part of these Registry standards.
VOResource describes an XML encoding standard for IVOA Resource Metadata. The primary intended use of VOResource is to provide an XML interchange format for use with resource registries. A registry is a repository of resource descriptions and is employed by users and applications to discover resources. VOResource can be used to pass descriptions from publishers to registries and then from registries to applications. Another intended use is as a language for services to describe themselves directly.

VOResource became an IVOA Recommended standard in February 2008.

## 4.3 Registry Interface

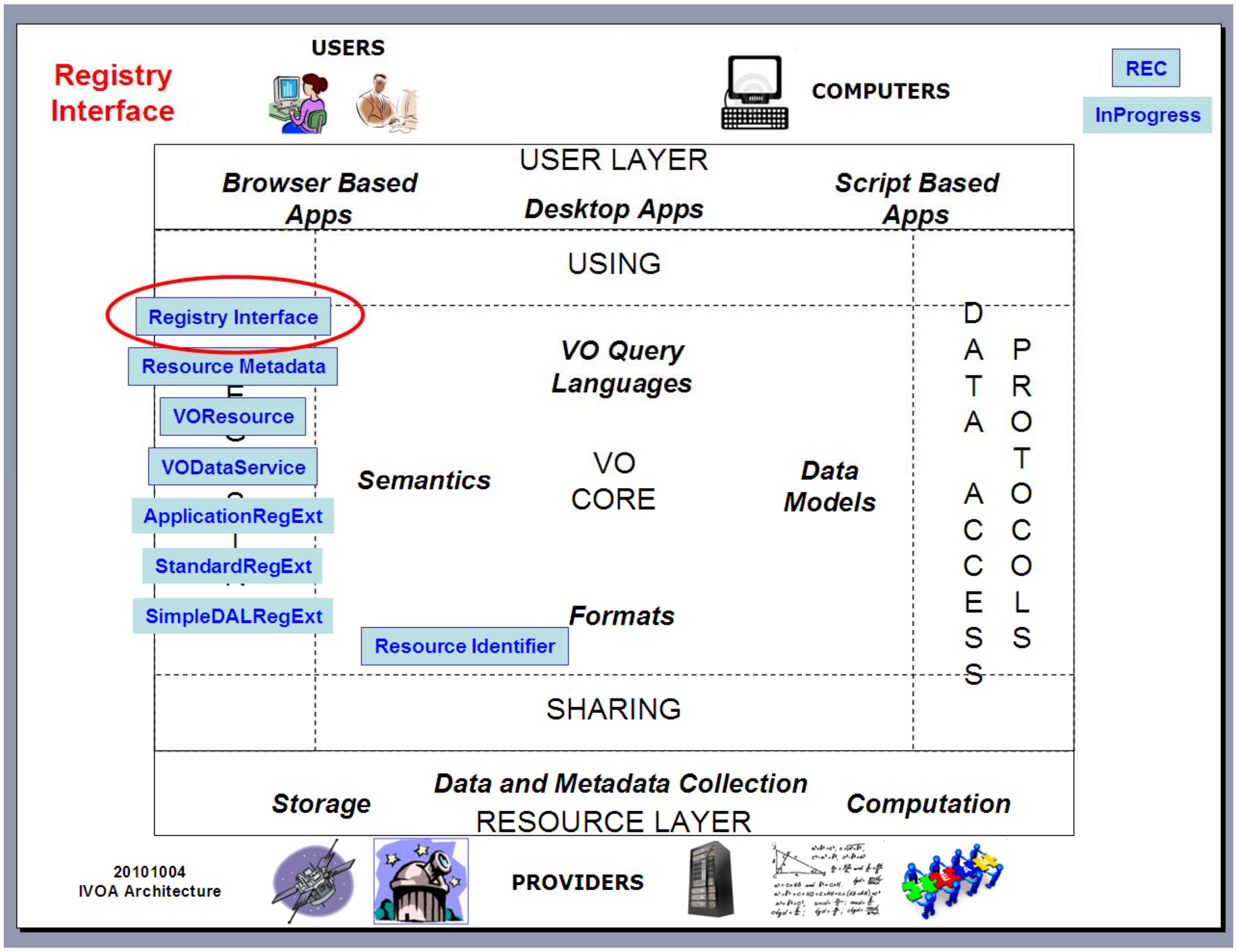

The IVOA Registry enables users and applications in the User Layer to discover data and metadata collection, as well services in the Resource Layer. The Registry Interface defines the interfaces that support interactions between applications and registries as well as between the registries themselves. It is based on a general, distributed model composed of so-called *searchable* and *publishing* registries. The specification has two main components: an interface for searching and an interface for *harvesting*. All interfaces are defined by a standard Web Service Description Language (WSDL) document; however, harvesting is also supported through the existing Open Archives Initiative Protocol for Metadata Harvesting, defined as an HTTP REST interface. Finally, Registry Interface details the metadata used to describe registries themselves as resources using an extension of the VOResource metadata schema.

Registry Interface makes reference to all existing registry standards, as they are all subject to being accessed through this interface, either by applications, or by other registries harvesting the associated resources.

Note that the Registry Interface standard makes use of an earlier version of ADQL (v1.01) which had never become an IVOA Recommendation. Hence, the Registry Interface includes an Annex containing the definition of this “ADQL 1.01” which does not correspond to the formal Recommended ADQL v2.0.

Registry Interface v1.0 became an IVOA Recommended standard in November 2009.

## 4.4 VODataService

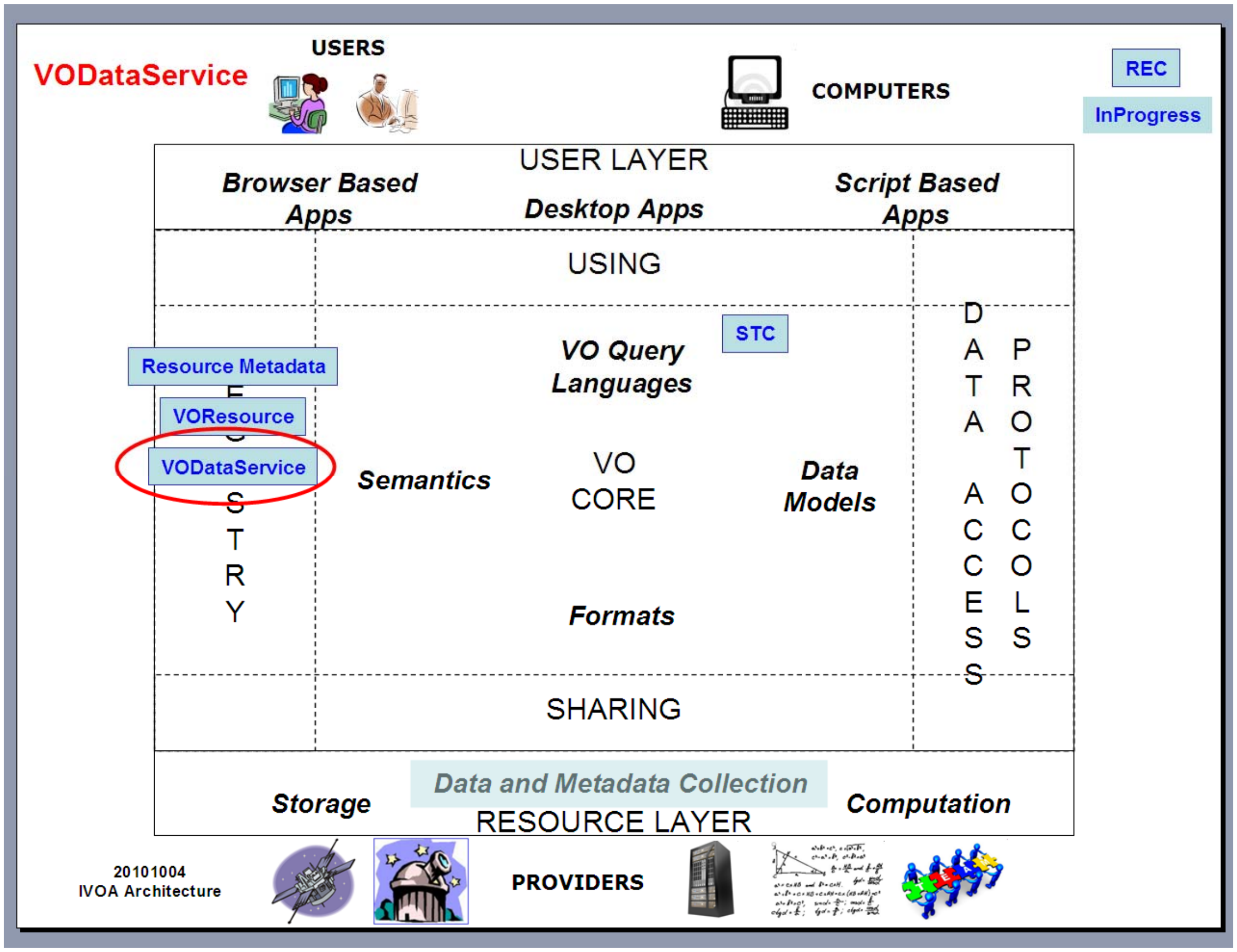

The IVOA Registry enables users and applications in the User Layer to discover data and metadata collection, as well services in the Resource Layer. VODataService standard is part of the registry standards which make this discovery possible.

VODataService refers to an XML encoding standard for a specialized extension of the IVOA Resource Metadata that is useful for describing data collections and the services that access them. It is defined as an extension of the core resource metadata encoding standard known as the VOResource standard using XML Schema. The specialized resource types defined by the VODataService schema allow one to describe how the data underlying the resource cover the sky as well as their frequency and time. This coverage description leverages heavily on the Space-Time Coordinates (STC) standard schema. VODataService also enables detailed descriptions of tables that include information useful to the discovery of tabular data.

VODataService 1.1 became an IVOA Recommended standard in October 2010.

## 4.5 ApplicationRegExt

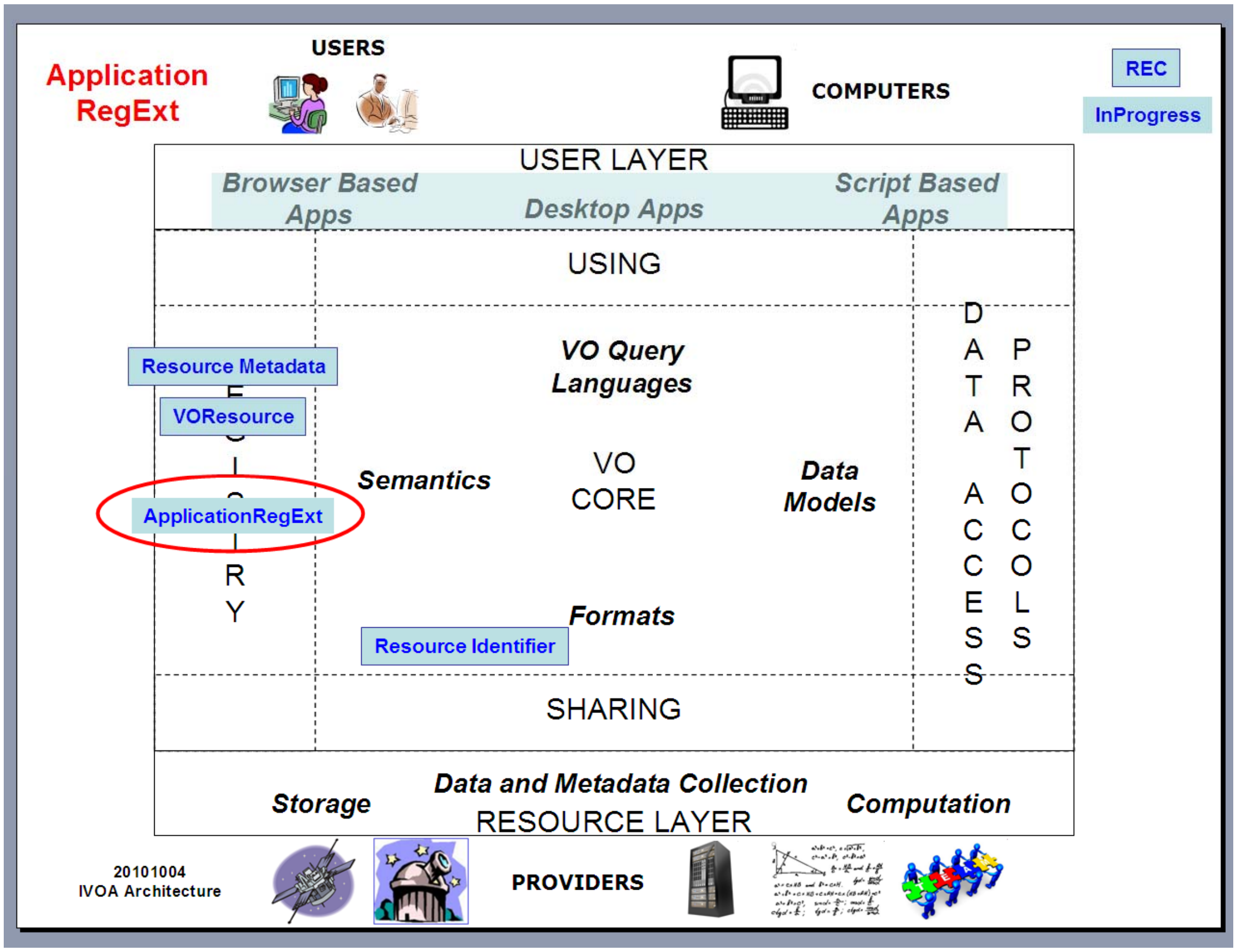

The IVOA Registry enables users and applications in the User Layer to discover data and metadata collection, as well services. Such a service could be a VO Application. ApplicationRegExt (for Applications Registry Extension) is part of these registry standards which make discovery VOApplications possible.

ApplicationRegExt refers to an XML encoding standard for a specialized extension of the IVOA Resource Metadata that is useful for describing VO Applications. It is defined as an extension the core resource metadata encoding standard known as the VOResource standard using XML Schema.

Detailed definition of ApplicationRegExt and its potential interdependencies with other IVOA standards still need to be defined. For example, one could imagine links to SAMP, SSO amongst others.

By registering a VO Application in a Registry, it gets a unique IVOA Resource Identifier which then can be referred to by other applications and services.

In 2010, registry efforts are going into other standards, so there is no planned date for the release of the ApplicationRegExt standard.

(Note that ApplicationRegExt was previously referred as “VOApplications”, but it was decided to change it to a more descriptive term).

## 4.6 StandardRegExt

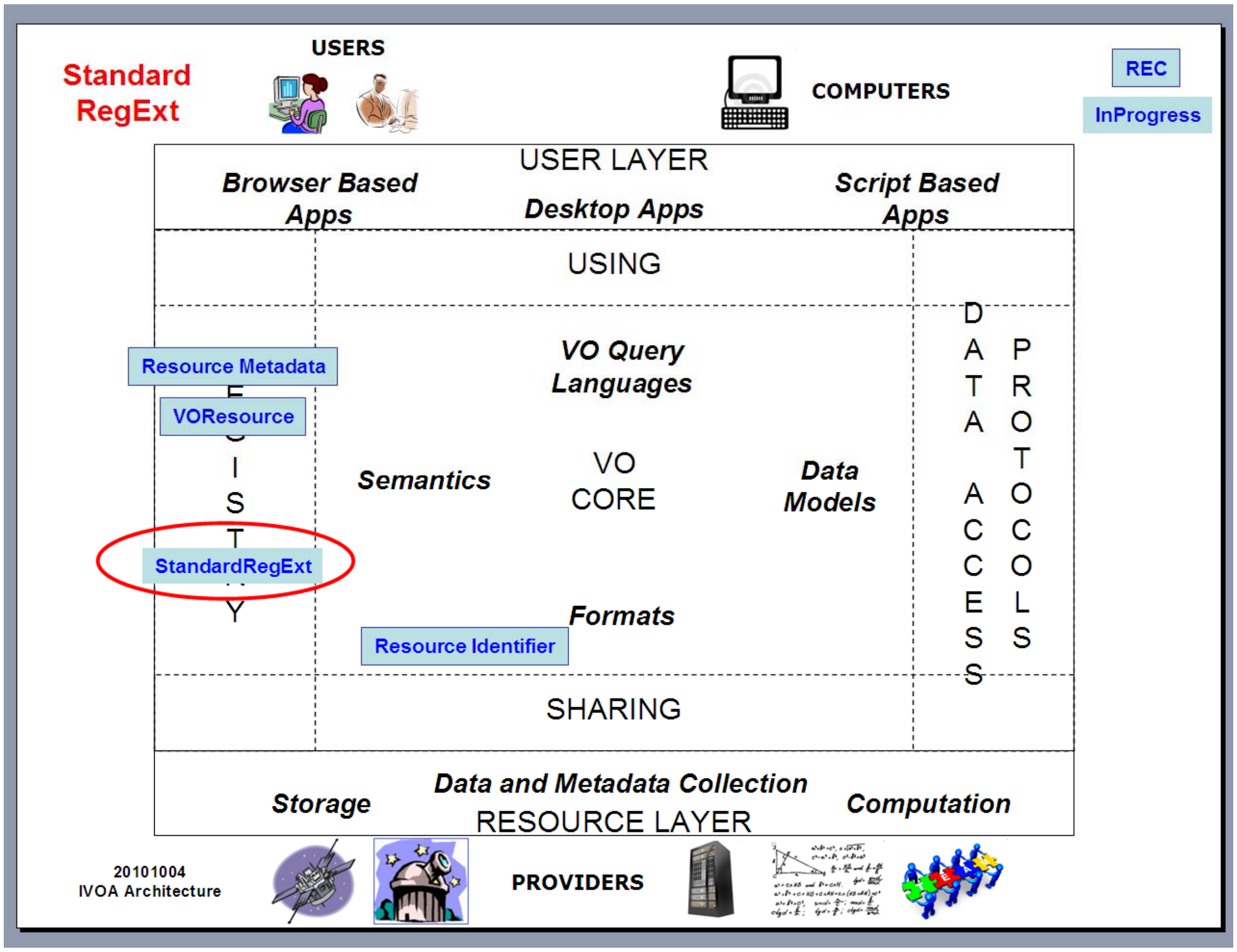

An important goal of the IVOA is to define and publish standards for data and computational services which can interoperate to create a Virtual Observatory (VO). Central to the coordination of these services is the concept of a registry where resources can be described and thus discovered by users and applications in the VO. StandardRegExt (for Standard Registry Extension) is part of these registry standards which defines an IVOA Standard.

StandardRegExt refers to an XML encoding standard for a specialized extension of the IVOA Resource Metadata that is useful for describing VO Standard. It is defined as an extension to the core resource metadata encoding standard known as the VOResource standard using XML Schema.

By registering an IVOA Standard in a Registry, it gets a unique IVOA Resource Identifier which then can be referred to in other resource descriptions, namely for services that support the standard.

In 2010, the StandardRegExt standard is still under development.

(Note that StandardRegExt was previously referred as “VOStandard”, but it was decided to change it to a more descriptive term).

## 4.7 SimpleDALRegExt

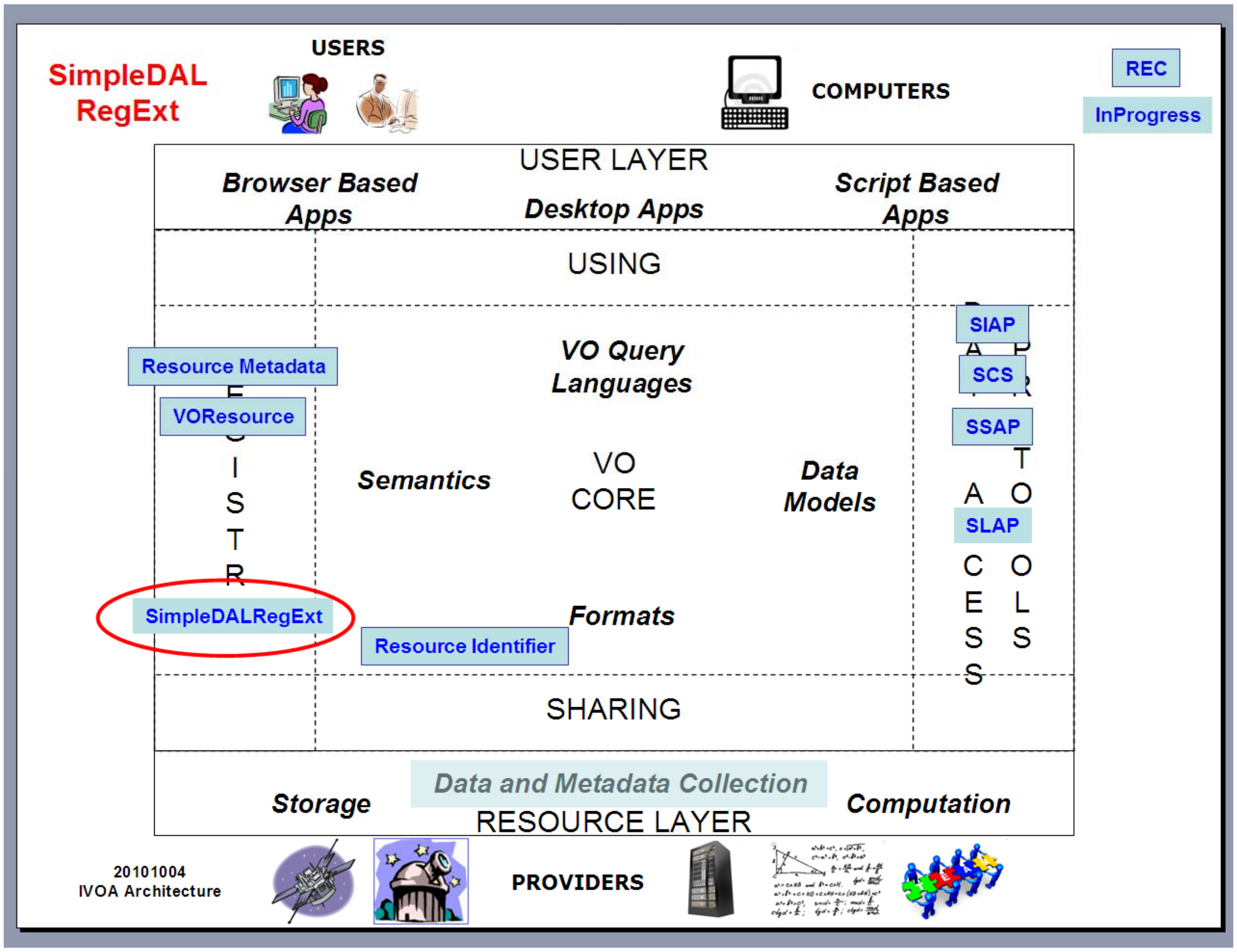

The IVOA Registry enables users and applications in the User Layer to discover data and metadata collection, as well services. Such archives could be data and metadata collections accessible through the IVOA Simple Access Protocols (ie SIAP, SCS, SSAP and SLAP). SimpleDALRegExt (for Simple Data Access Layer Protocols Registry Extension) is part of these registry standards which make discovery of Simple DAL services possible.

SimpleDALRegExt refers to an XML encoding standard for a specialized extension of the IVOA Resource Metadata that is useful for describing VO Simple DAL Services. It is defined as an extension the core resource metadata encoding standard known as the VOResource standard using XML Schema.

Detailed definition of SimpleDALRegExt and its potential interdependencies with other IVOA standards still need to be defined. For example, one could imagine interdependencies with SEAP, SimDAL, FAP, etc...

In 2010, the SimpleDALRegExt standard is still under development.

(Note that SimpleDALRegExt was previously referred as “RegSimpleDAL”, but it was decided to change it to a more descriptive term).

## 4.8 VOSI – VO Support Interface

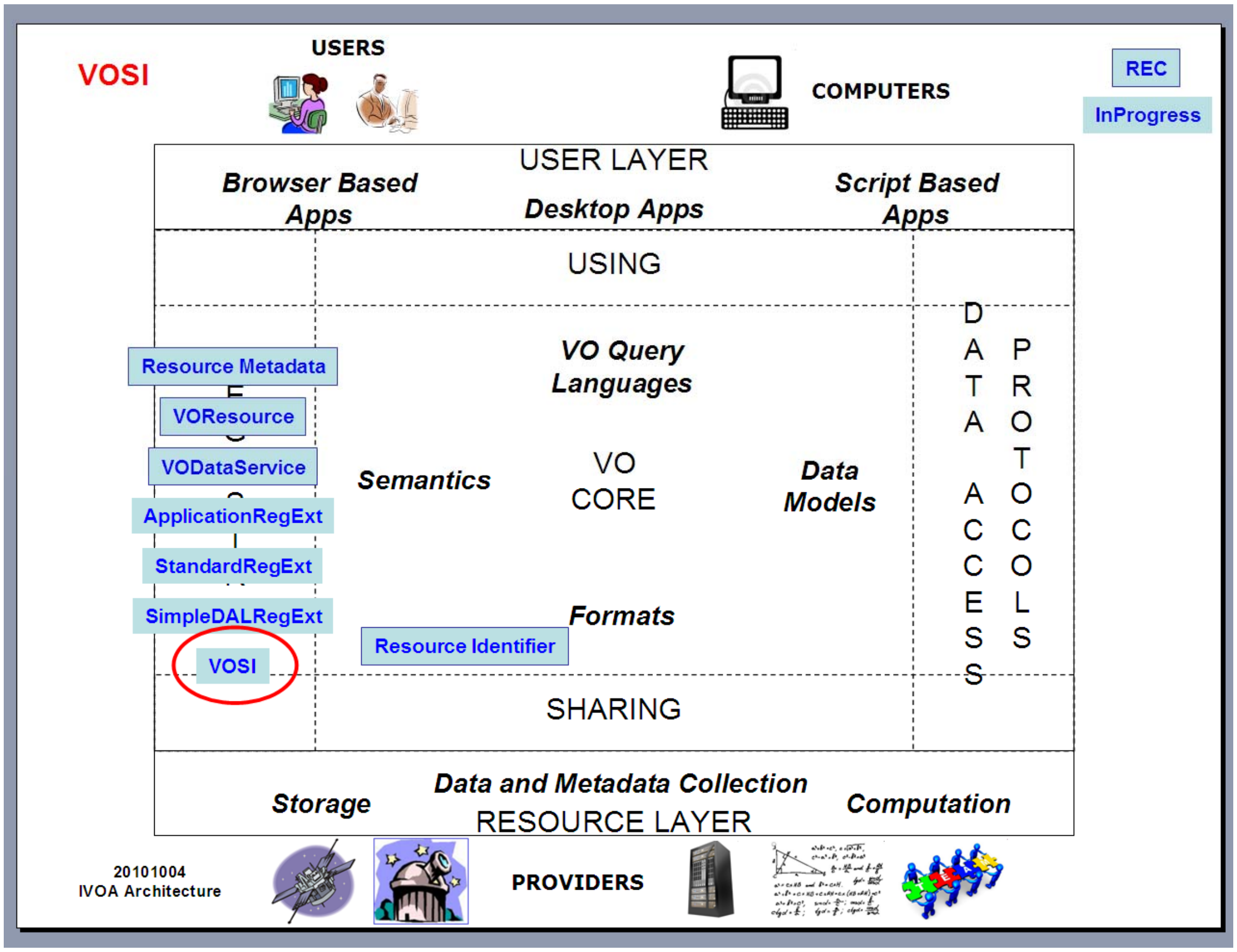

Much of the VO is made up of web services which are registered into the VO Registry and get a unique Resource Identifier. VOSI (standing for VO Support Interface) is the standard that defines the basic functions that all VO services should provide in order to support management of the VO.
For example, it describes the capabilities of any VO Resource (which IVOA standard function is being support (for example an SSAP + SSLDM service) and how to invoke this service, including records of any details of the implementation of the function that are not fixed in the standard for that function. VOSI also defines the various possible states of this service (such as available, up since, down at, back at, etc ...).

VO Support Interfaces is expected to become an IVOA Recommended standard in 2010.

# 5 Getting / Data Access Protocol Standards

Astronomical data and metadata collections are distributed and being made available to the scientific community through dedicated access mechanisms. This is usually done through the internet, via web interfaces, FTP services and other web services. These access mechanisms differ between various data and metadata collections and from one provider to another.

IVOA Data Access Protocols provide standardized mechanisms for querying and access to distributed data and metadata collections. They allow users and applications to access in a consistent and uniform way datasets stored in different locations that were previously available only through dedicated interfaces.

Data Access Protocols can be “simple”, i.e. almost self contained (although usually making reference to some of the VO Core standards) and offering simple query and access mechanism. Initial IVOA Data Access Protocols were created in this context, allowing faster implementation and take-up, acknowledging their inherent limitations due to their “simplicity”.

With the adoption of VO Query Languages, Data Models and the Table Access Protocol Standard, more powerful query capabilities are enabled for discovery and access to data.

Data Access Protocols services are to be registered into a VO Registry according to the associated registry resources standards.

## 5.1 SIAP – Simple Image Access Protocol

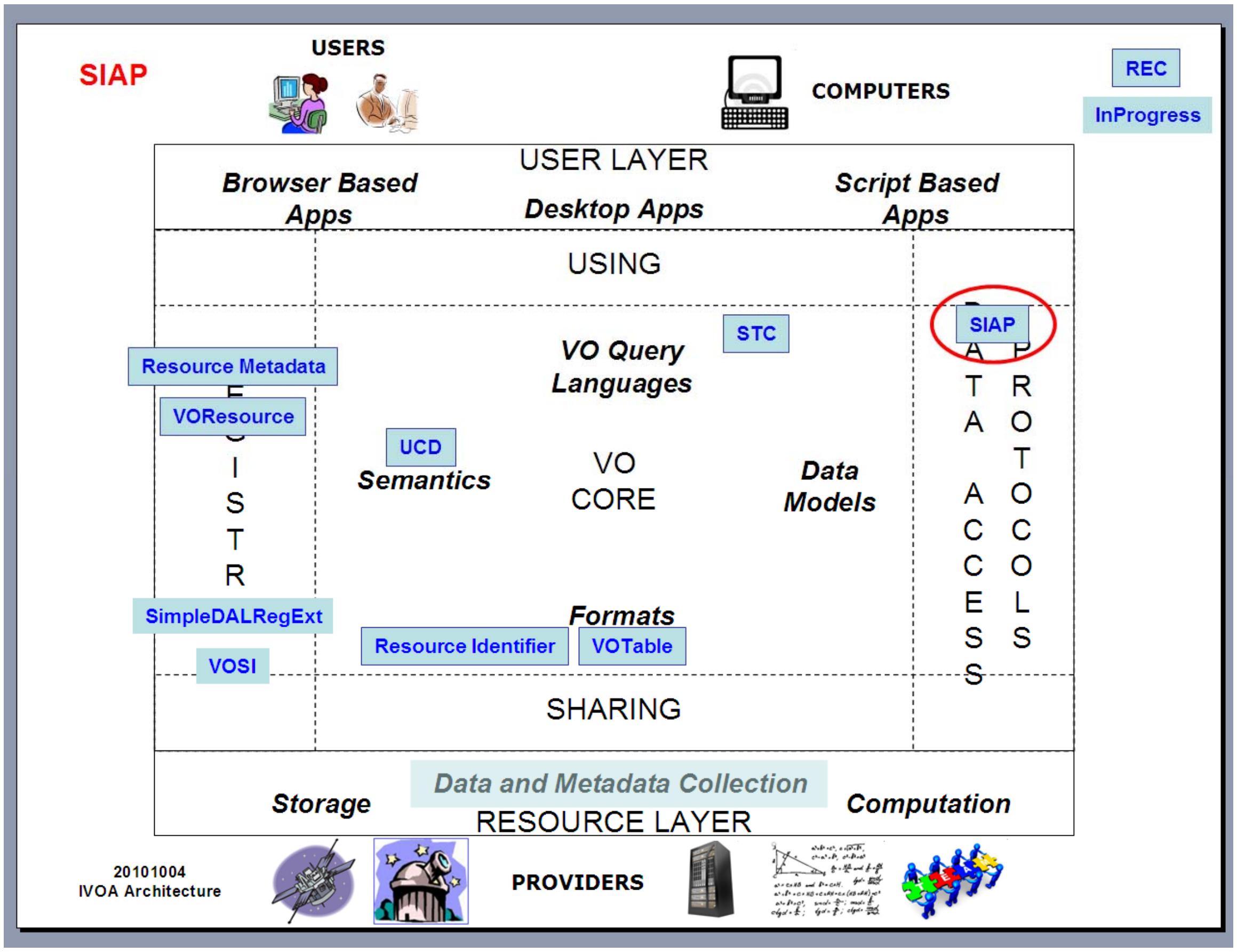

IVOA Data Access Protocols provide standardized mechanisms for querying and access to distributed data and metadata collections. They allow users and applications to access in a consistent and uniform way datasets stored in different locations that were previously available only through dedicated interfaces.

Simple Image Access Protocol (SIAP) defines an interface for querying and retrieving image data from a variety of astronomical image repositories through a uniform interface. The interface is meant to be reasonably simple to implement by service providers. A query defining a rectangular region on the sky is used to query for candidate images. The service returns a list of candidate images formatted as a VOTable. For each candidate image an access reference URL may be used to retrieve the image. Images may be returned in a variety of formats including FITS and various graphics formats.

As with most of the IVOA Data Access Protocol, SIAP makes use of VOTable for metadata exchange, STC and UCD for metadata description. However, as one of the very early IVOA standard, SIAP does not use Utypes and Units for metadata description.

A SIAP service is to be registered into the VO Registry, using the associated registry standards (in particular Resource Metadata, VOResource and SimpleDALRegExt specifically for Data Access Protocols such as SIAP). Once registered, a SIAP service will get a unique IVOA Resource Identifier. Furthermore, a SIAP service should be registered with its supported interfaces through the VOSI standard.

SIAP v1.0 became an IVOA Recommended standard in November 2009.

## 5.2 SCS – Simple Cone Search

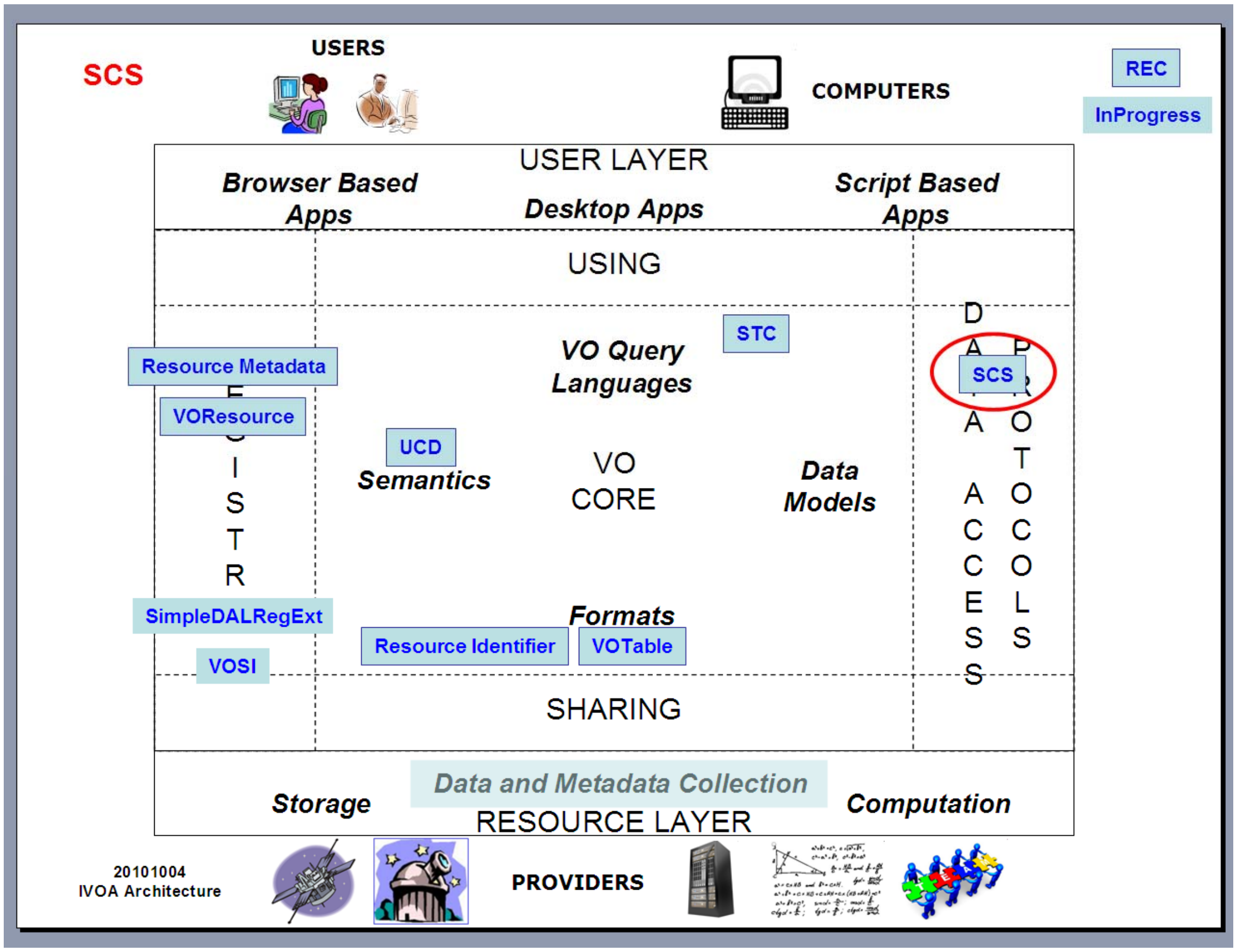

IVOA Data Access Protocols provide standardized mechanisms for querying and access to distributed data and metadata collections. They allow users and applications to access in a consistent and uniform way datasets stored in different locations that were previously available only through dedicated interfaces.

Simple Cone Search defines a simple query protocol for retrieving records from a catalog of astronomical sources. The query describes sky position and an angular distance, defining a cone on the sky. The response returns a list of astronomical sources from the catalog whose positions lie within the cone, formatted as a VOTable.

As with most of the IVOA Data Access Protocol, SCS makes use of VOTable for metadata exchange, STC and UCD for metadata description. However, as one of the very early IVOA standard, SCS does not use Utypes and Units for metadata description.

An SCS service is to be registered into the VO Registry, using the associated registry standards (in particular Resource Metadata, VOResource and SimpleDALRegExt specifically for Data Access Protocols such as SCS). Once registered, an SCS service will get a unique IVOA Resource Identifier. Furthermore, an SCS service should be registered with its supported interfaces through the VOSI standard.

SCS v1.03 became an IVOA Recommended standard in February 2008.

## 5.3 SSAP – Simple Spectra Access Protocol

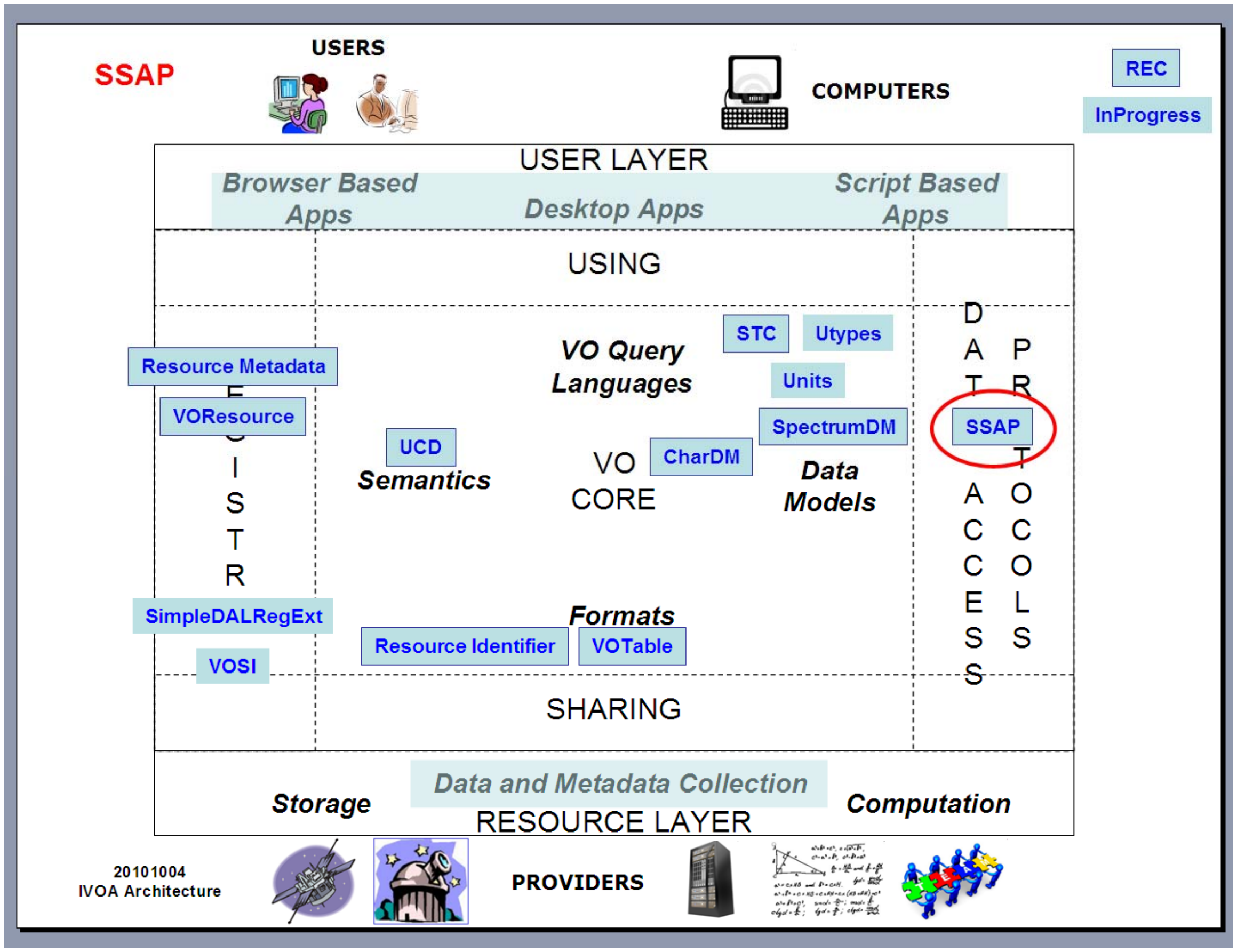

IVOA Data Access Protocols provide standardized mechanisms for querying and access to distributed data and metadata collections. They allow users and applications to access in a consistent and uniform way datasets stored in different locations that were previously available only through dedicated interfaces.

Simple Spectra Access Protocol (SSAP) defines a uniform interface to remotely discover and access one dimensional spectrum coming from data and metadata collections. SSAP is based on the Spectrum Data Model (itself making reference to the CharDM) that is capable of describing most tabular spectrophotometric data, including time series and spectral energy distributions (SEDs) as well as 1-D spectra. SSAP can be used from VO applications to access the associated spectrum resources.

As with most of the IVOA Data Access Protocols, SSAP makes use of VOTable for metadata exchange, STC, UCD, Utypes and Units for metadata description.

An SSAP service is to be registered into the VO Registry, using the associated registry standards (in particular Resource Metadata, VOResource and SimpleDALRegExt specifically for Data Access Protocols such as SSAP). Once registered, an SSAP service will get a unique IVOA Resource Identifier. Furthermore, an SSAP service should be registered with its supported interfaces through the VOSI standard.

SSAP v1.04 became an IVOA Recommended standard in February 2008.

## 5.4 SLAP – Simple Line Access Protocol

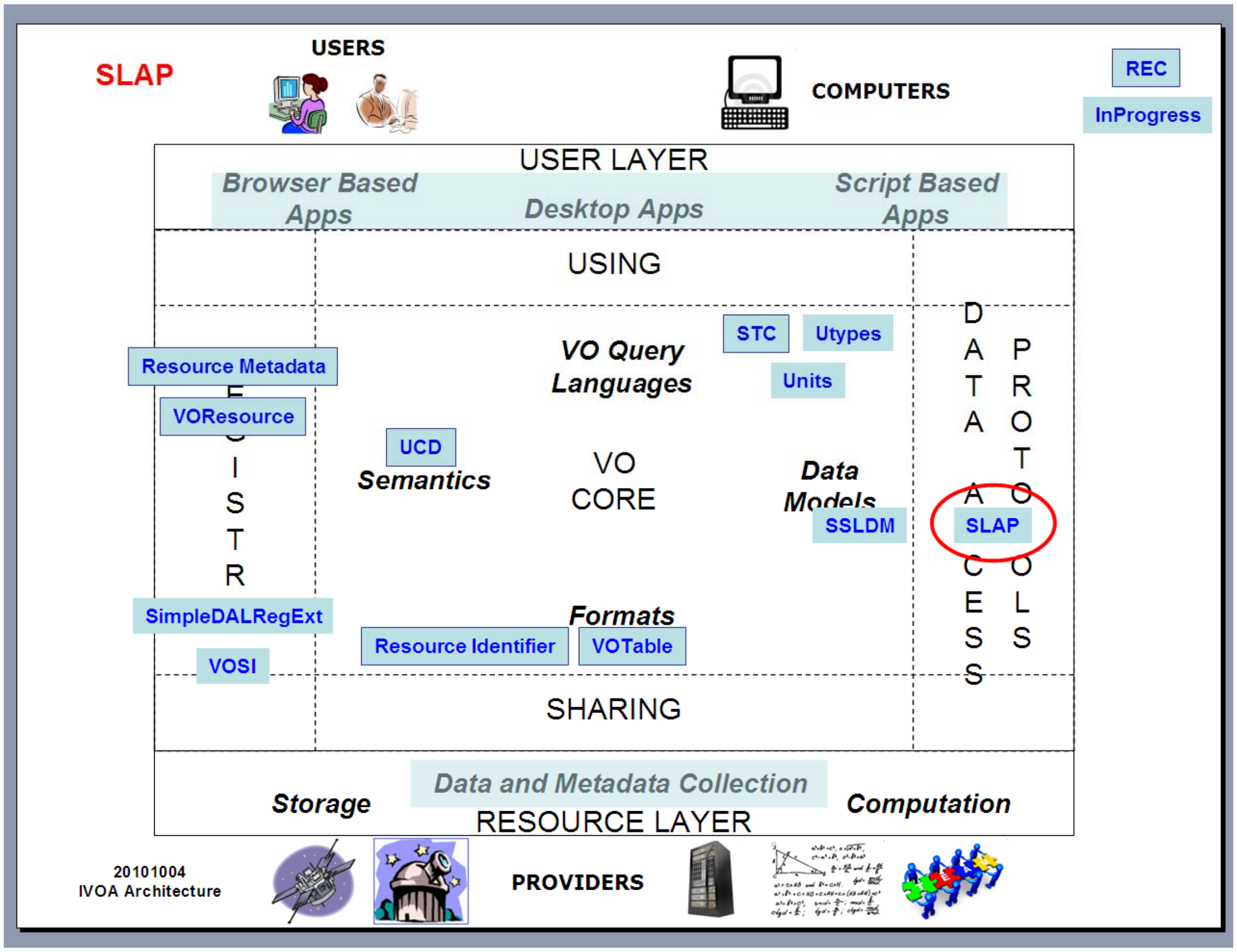

IVOA Data Access Protocols provide standardized mechanisms for querying and access to distributed data and metadata collections. They allow users and applications to access in a consistent and uniform way datasets stored in different locations that were previously available only through dedicated interfaces.

Simple Line Access Protocol (SLAP) defines a uniform interface to discover and access spectral line transitions coming from data and metadata collections. SLAP is based on the SSLDM (Simple Spectral Line Data Model).

As with most of the IVOA Data Access Protocols, SLAP makes use of VOTable for metadata exchange, STC, UCD, Utypes and Units for metadata description.

A SLAP service is to be registered into the VO Registry, using the associated registry standards (in particular Resource Metadata, VOResource and SimpleDALRegExt specifically for Data Access Protocols such as SLAP). Once registered, a SLAP service will get a unique IVOA Resource Identifier.

Furthermore, SLAP services should be registered with its supported interfaces via the VOSI standard.

SLAP is currently under review and it expected to become an IVOA Recommended standard in 2010.

## 5.5 TAP – Table Access Protocol

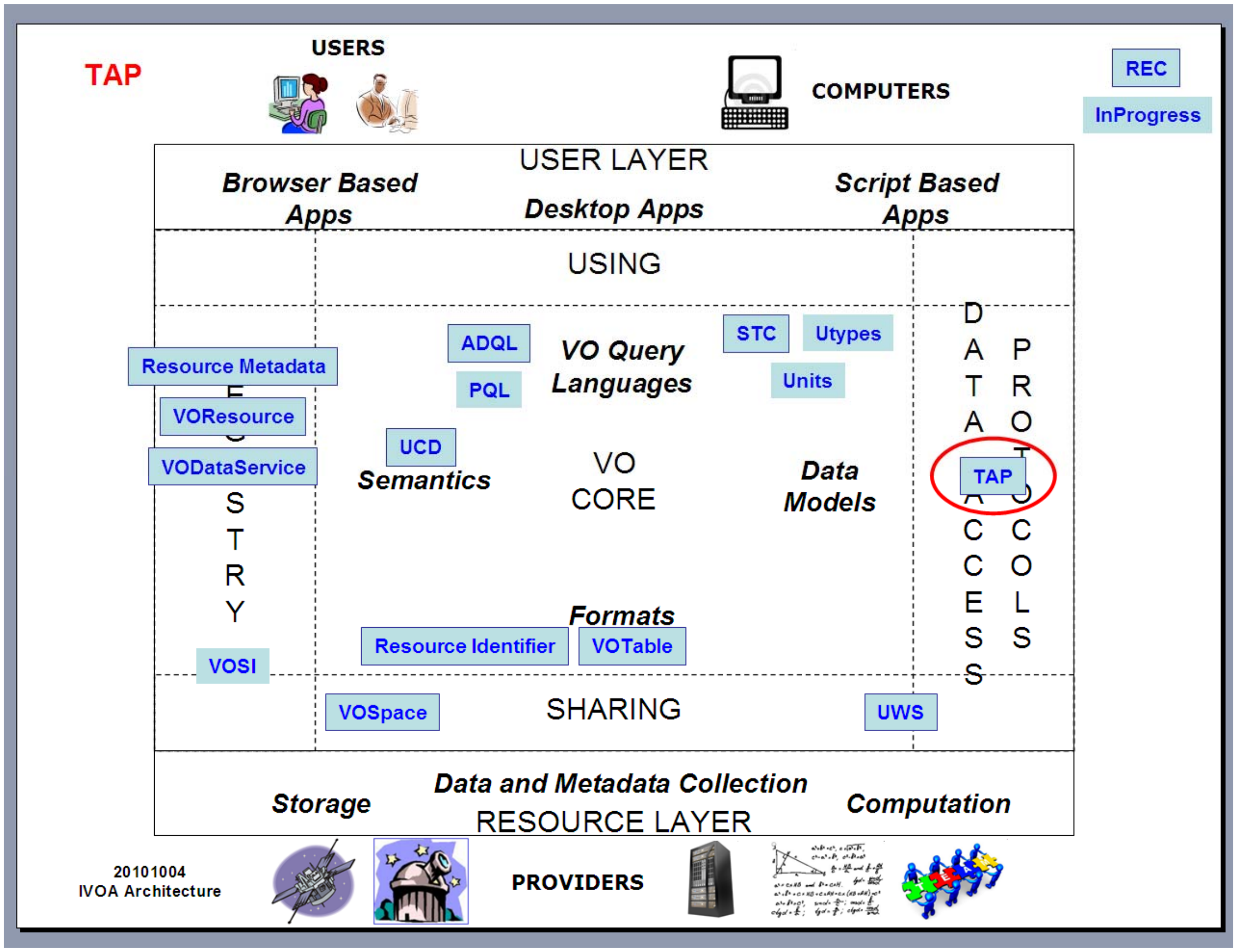

IVOA Data Access Protocols provide standardized mechanisms for querying and access to distributed data and metadata collections. They allow users and applications to access in a consistent and uniform way datasets stored in different locations that were previously available only through dedicated interfaces.

The Table Access Protocol (TAP) defines a uniform interface for querying general table data, including astronomical catalogs as well as general database tables. Access is provided for both database and table metadata as well as for actual table data. This version of the protocol includes support for multiple query languages, including the Astronomical Data Query Language (ADQL) and the Parameterized Query Language (PQL). It also includes support for both synchronous and asynchronous queries.

TAP services support three kinds of queries: data queries, metadata queries, and Virtual Observatory Support Interface (VOSI) queries.

Asynchronous query support is mandatory in TAP. Asynchronous queries require that client and server share knowledge of the state of the query during its execution and between HTTP exchanges. In TAP, the mechanism by which the clients and services share the state of transactions is based on the Universal Worker Service (UWS) pattern.

As with most of the IVOA Data Access Protocols, TAP makes use of VOTable for metadata exchange, STC, UCD, Utypes and Units for metadata description.

This version of TAP is currently making a limited use of VOSpace, but further support is planned for future TAP versions.

A TAP service is to be registered into the VO Registry, using the associated registry standards (in particular Resource Metadata, VOResource and VODataResource). Once registered, a TAP service will get a unique IVOA Resource Identifier. Furthermore, TAP services should be registered with its supported interfaces via the VOSI standard.

TAP v1.0 became an IVOA Recommended standard in March 2010.

## 5.6 SEAP – Simple Event Access Protocol

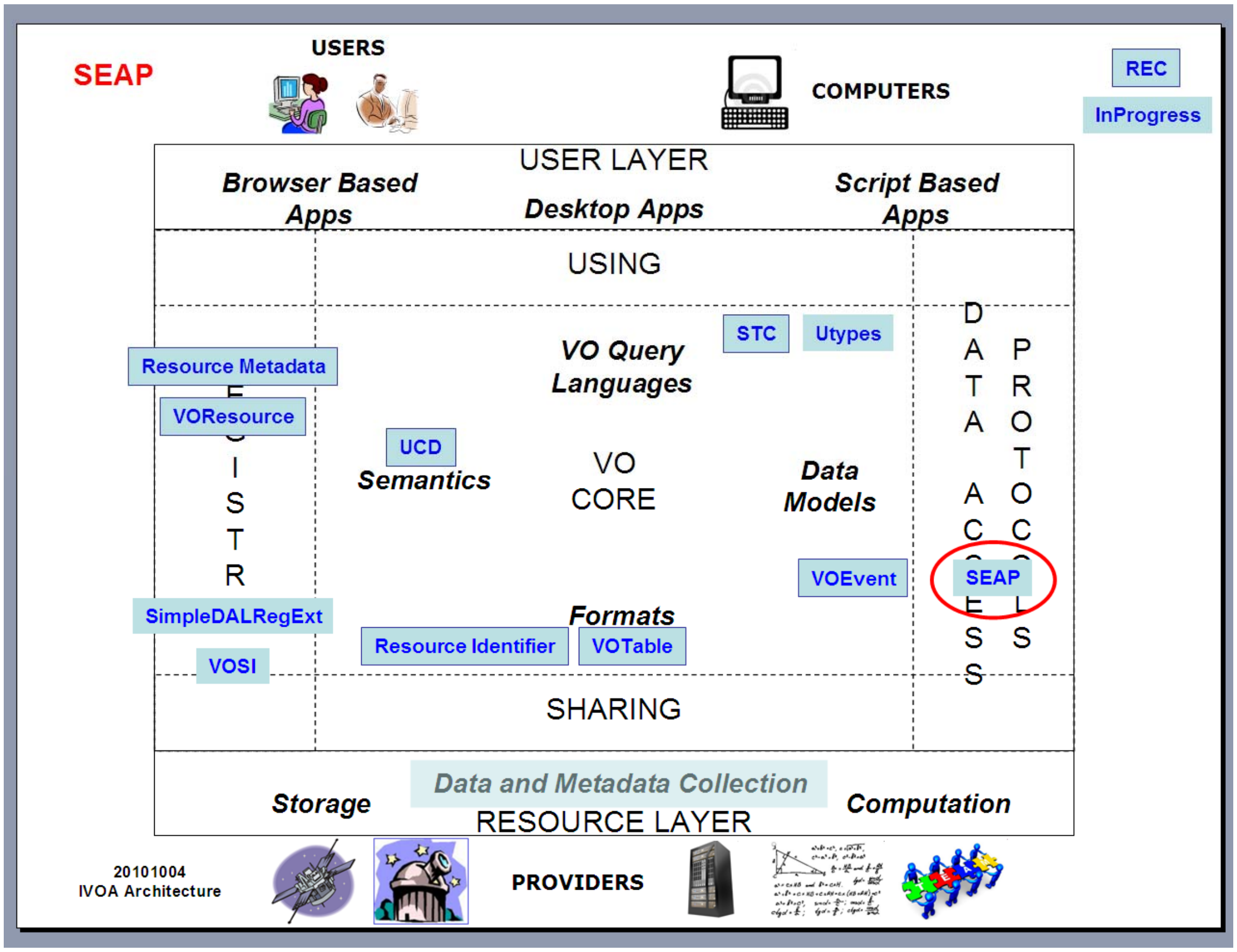

IVOA Data Access Protocols provide standardized mechanisms for querying and access to distributed data and metadata collections. They allow users and applications to access in a consistent and uniform way datasets stored in different locations that were previously available only through dedicated interfaces.

Simple Event Access Protocol (SEAP) defines a uniform interface to remotely discover and access astronomical Events coming from data and metadata collections. SEAP is based on the VOEvent Data Model that defines the content and meaning of a standard information packet for representing, transmitting, publishing and archiving the discovery of a transient celestial event, with the implication that timely follow-up is being requested.

In 2010, the SEAP standard is still under development. Its relationship with other IVOA standards is expected to be similar to other Simple DAL protocols, but this still needs to be confirmed.

As with most of the IVOA Data Access Protocols, SEAP is expected to use of VOTable for metadata exchange, STC, UCD, Utypes and Units for metadata description.

A SEAP service is to be registered into the VO Registry, using the associated registry standards (in particular Resource Metadata, VOResource and SimpleDALRegExt specifically for Data Access Protocols such as SEAP). Once registered, a SEAP service will get a unique IVOA Resource Identifier. Furthermore, SEAP services should be registered with its supported interfaces via the VOSI standard.

## 5.7 SimDAL – Simulations Data Access Layer

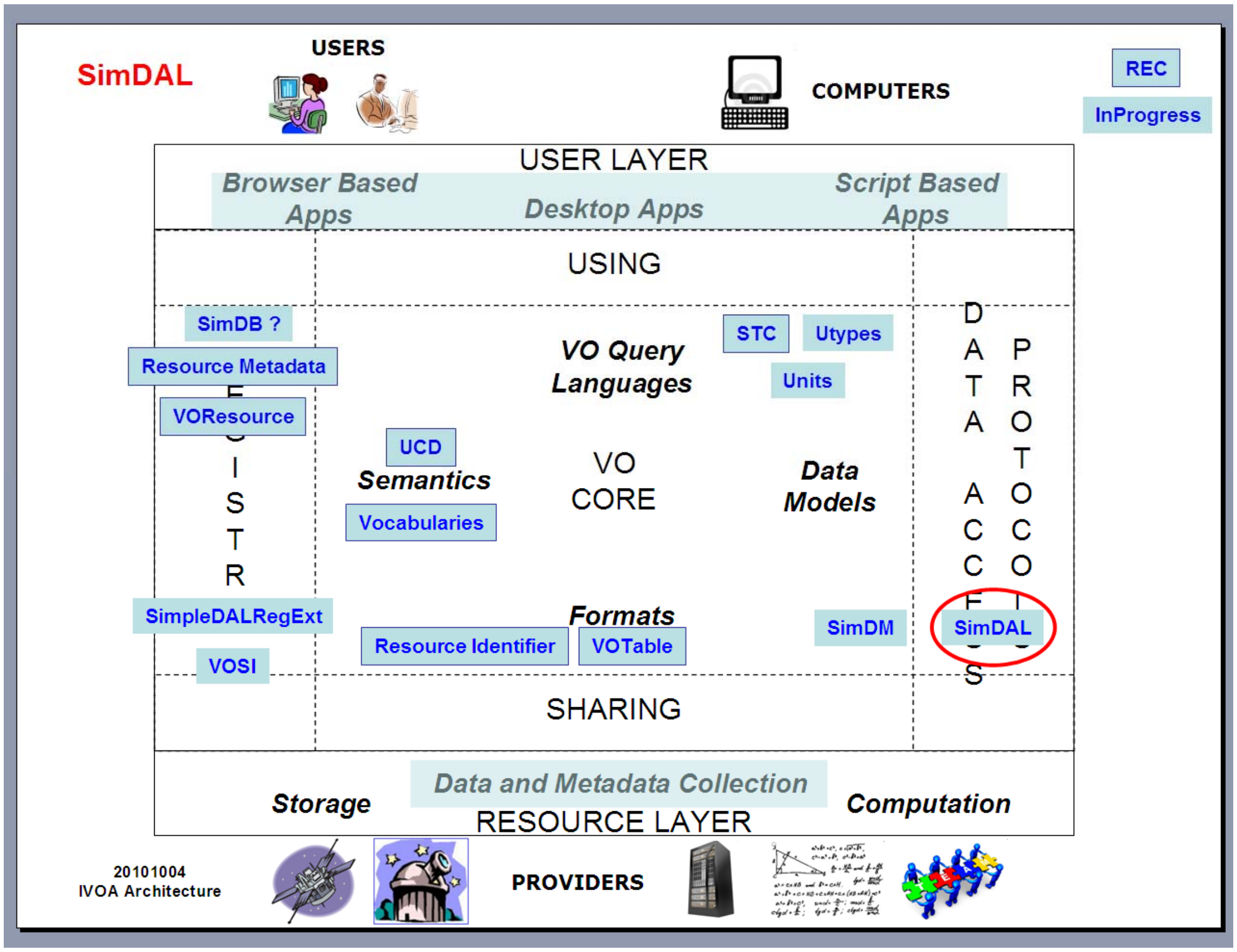

IVOA Data Access Protocols provide standardized mechanisms for querying and access to distributed data and metadata collections. They allow users and applications to access in a consistent and uniform way datasets stored in different locations that were previously available only through dedicated interfaces.

The Simulation Data Access Layer (SimDAL) defines a uniform way to access simulation and theoretical products, working in conjunction with the Simulation Data Model (SimDM). SimDAL now encompasses in a single standard the previously named SimDAP and S3 draft standards.

In 2010, the SimDAL standard is still under development. The detailed definition of SimDAL and its potential interdependencies with other IVOA standards still need to be defined, so the above diagram is based on similar diagrams from other VO Data Access Protocols.

As with most of the IVOA Data Access Protocols, SimDAL will probably make use of VOTable for metadata exchange, UCD, Utypes and Units for metadata description. It might make use of a dedicated Vocabulary for simulation datasets.

Details about the registration still need to be clarified, whether SimDAL will be registered in the VO Registry

- like other DAL services (through the use of the associated registry standards (in particular Resource Metadata, VOResource and SimpleDALRegExt specifically for Data Access Protocols such as SimDAL), getting a unique IVOA Resource Identifier and supporting the VOSI standard).
- or through a dedicated Simulation DataBase (SimDB), including the definition of the interface to a database containing meta data describing simulations

## 5.8 FAP – Footprint Access Protocol

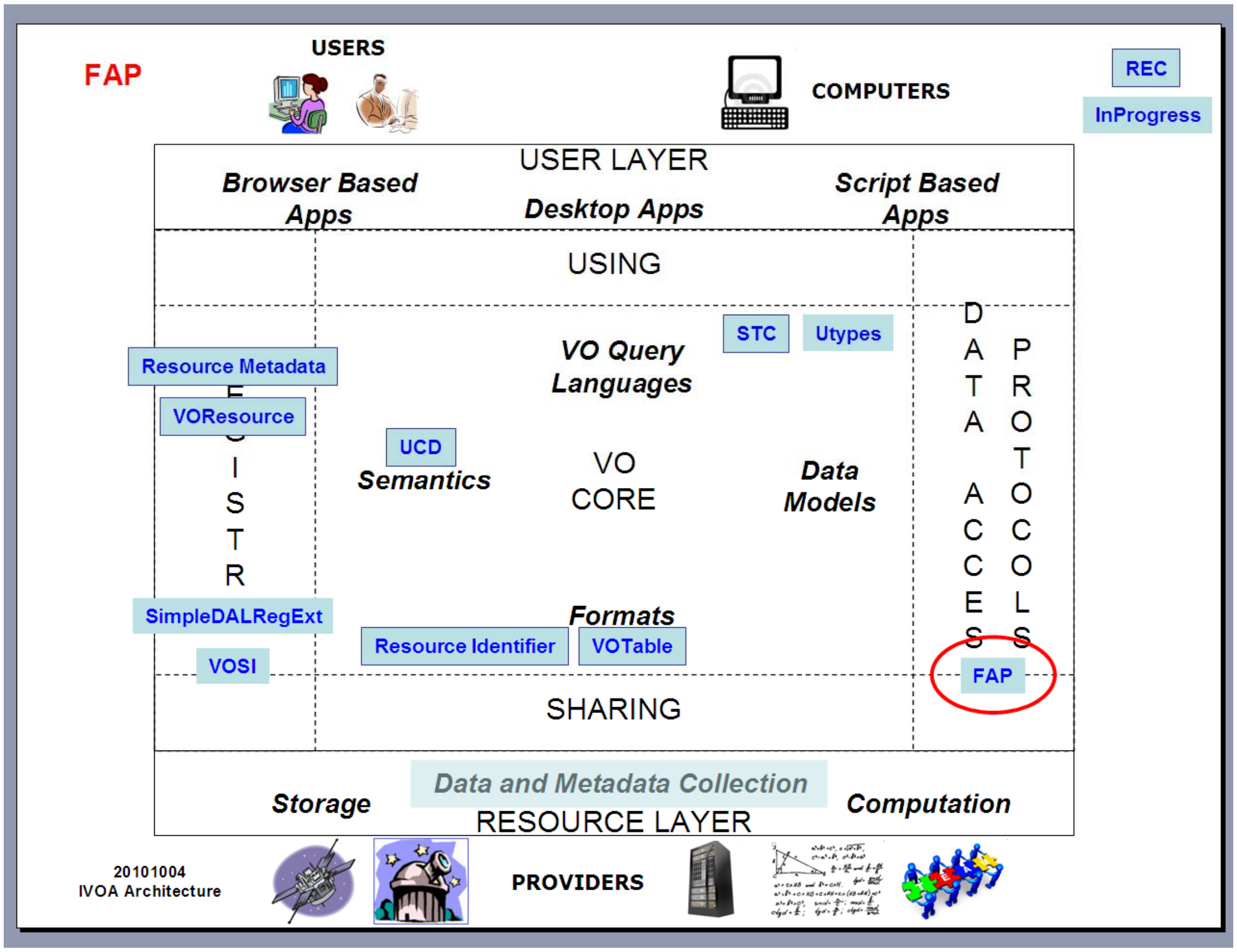

IVOA Data Access Protocols provide standardized mechanisms for querying and access to distributed data and metadata collections. They allow users and applications to access in a consistent and uniform way datasets stored in different locations that were previously available only through dedicated interfaces.

A footprint is a polygonal description of the spatial extent of a dataset. The Footprint Access Protocol (FAP) defines a uniform interface to remotely discover and access footprints coming from data and metadata collections.

In 2010, the FAP standard is still under development. Its relationship with other IVOA standard is expected to be similar to other Simple DAL protocols, but this still needs to be confirmed.

As with most of the IVOA Data Access Protocols, FAP is expected to use of VOTable for metadata exchange, STC, UCD, Utypes and Units for metadata description.

A FAP service is to be registered into the VO Registry, using the associated registry standards (in particular Resource Metadata, VOResource and SimpleDALRegExt specifically for Data Access Protocols such as FAP). Once registered, a FAP service will get a unique IVOA Resource Identifier. Furthermore, FAP services should be registered with its supported interfaces via the VOSI standard.

# 6 VO Core: Data Models Standards

Although common formats might be used (ie FITS), data providers usually represent and store their data and metadata collection according to their own needs. This representation is usually different from one provider to the other and can even be different for different data and metadata collections from the same provider.

Data Models in the VO aim to define the common elements of astronomical data and metadata collections and to provide a framework for describing their relationships so these become interoperable in a transparent manner. Complex query and access mechanisms can then be built with a VO query language (e.g. ADQL) onto different data and metadata collections that are published using common VO data models.

Some of the VO Data Models are specific to a type of data collection (e.g. Spectrum DM, SSLDM, ObsCoreDM, ObsProvDM, PhotDM, SimDM), while others (e.g. STC, Units, Utypes, CharDM) are more foundational and are components of the more specific Data Models. In general, each Data Model can be considered as a building block which can be referred from some other Data Models.

## 6.1 STC – Space Time Coordinate metadata

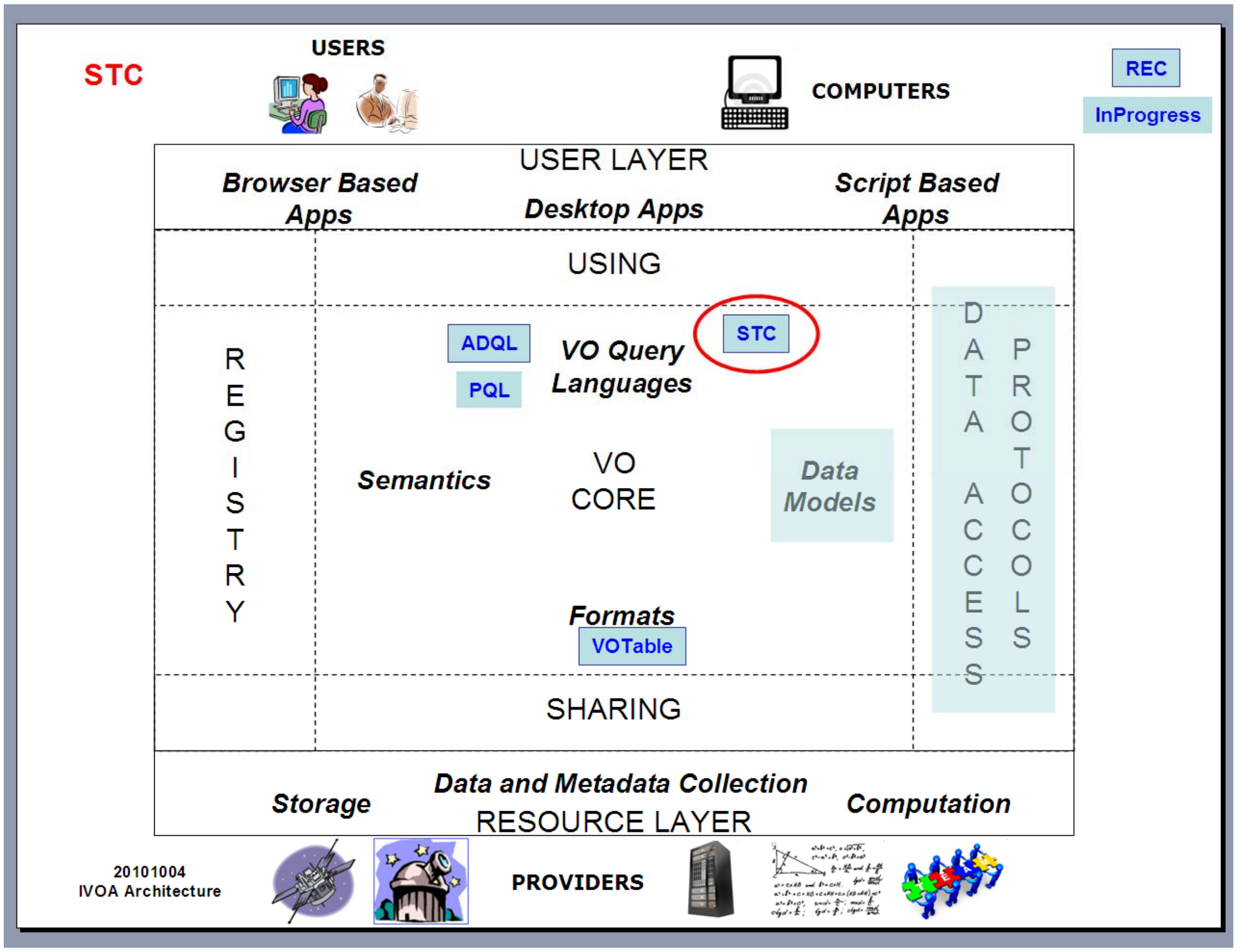

Data Models in the VO aim to define the common elements of astronomical data and metadata collections and to provide a framework for describing their relationships so these become inter operable in a transparent manner.

The Space-Time Coordinate (STC) standard provides a framework for describing spatial and temporal metadata. STC is used by most of the IVOA standards that refer to spatial and/or temporal metadata, e.g. IVOA Data Model standards, Data Access Protocol standards and VO Query Languages. STC can be serialized with a VOTable

STC v1.33 became an IVOA Recommended standard in October 2007.

## 6.2 Units

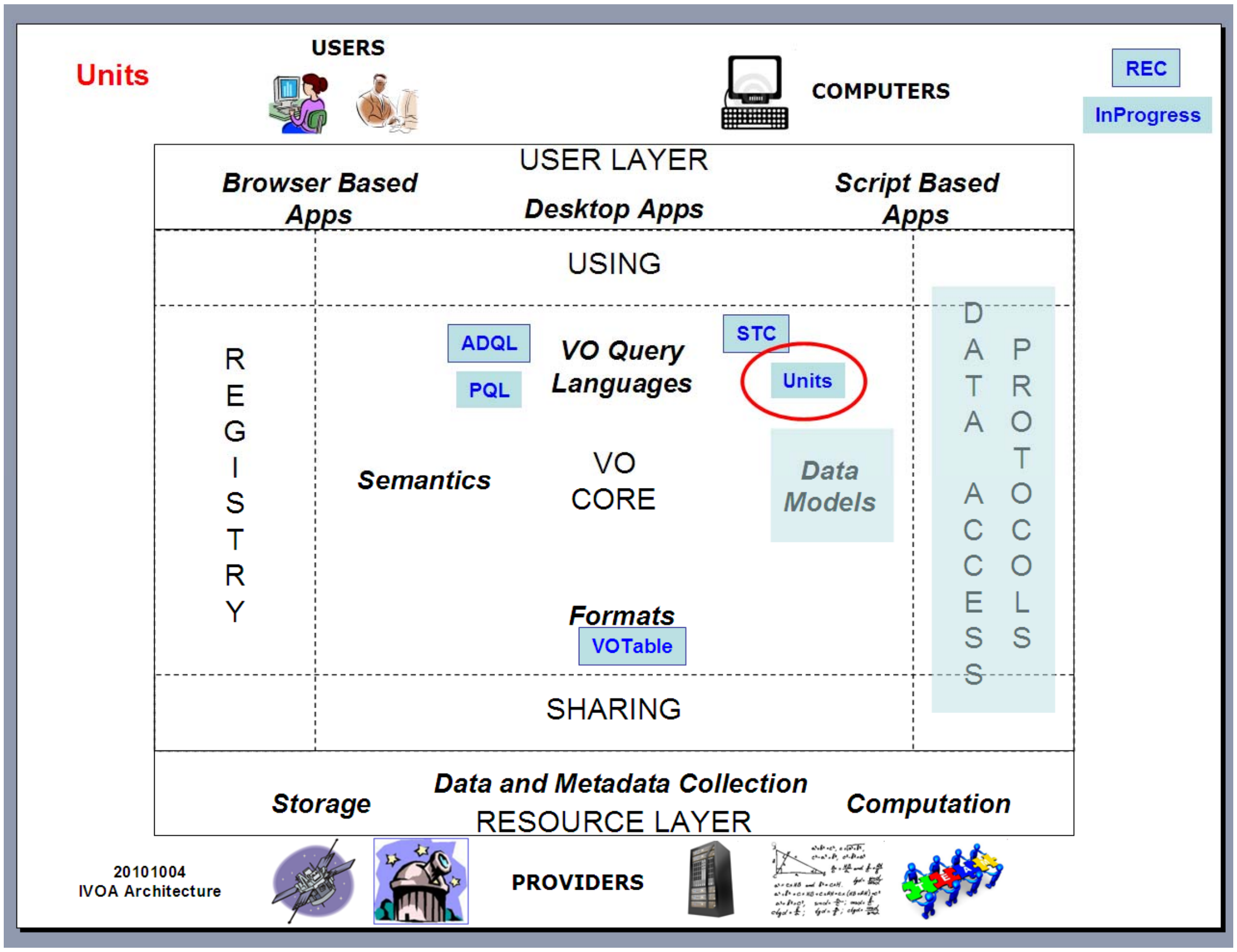

Data Models in the VO aim to define the common elements of astronomical data and metadata collections and to provide a framework for describing their relationships so these become inter operable in a transparent manner. This is particularly true for units usage, where there are rich and overlapping naming convention and representations. The VO Units DM aims at defining common practices in manipulating units in astronomical metadata and will define a means of consistent representation within VO services.

Detailed definition of Units DM and its potential interdependencies with other IVOA standards still need to be defined.

In 2010, the Units DM standard is still under development.

## 6.3 Utypes

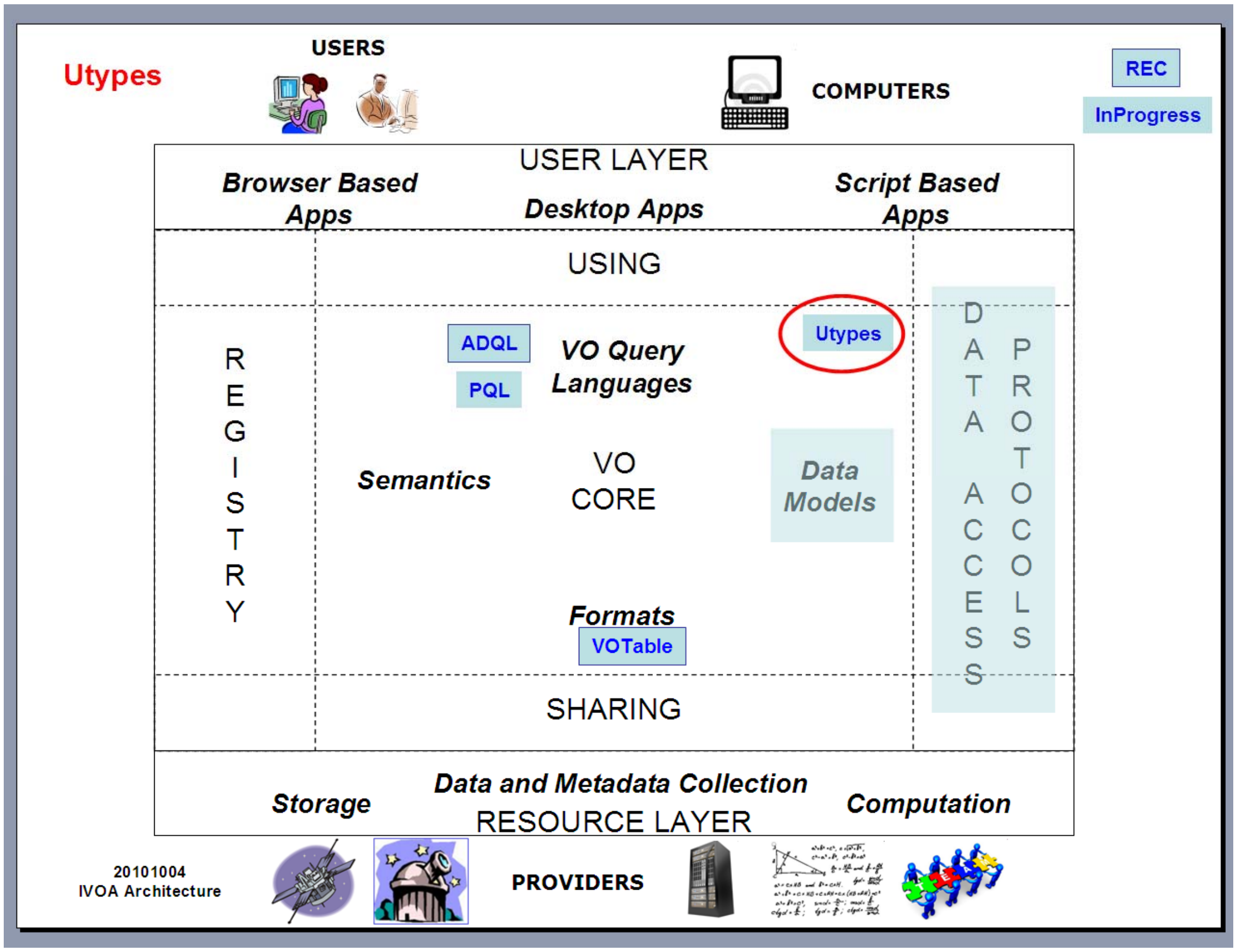

Data Models in the VO aim to define the common elements of astronomical data and metadata collections and to provide a framework for describing their relationships so these become inter operable in a transparent manner. A Data Model is composed of various elements representing some specific metadata. Utypes are names that define unambiguously an element of a data model that represent a piece of metadata in the VO.

Utypes DM will be used by all IVOA Data Models, as well in the context of Data Access Protocols and VO Query Languages. It will be naturally linked to the Units DM when available, and be used for metadata interchange through VOTable.

Detailed definition of Utypes DM and its potential interdependencies with other IVOA standards still need to be defined.

In 2010, the Utypes DM standard is still under development.

## 6.4 SpectrumDM

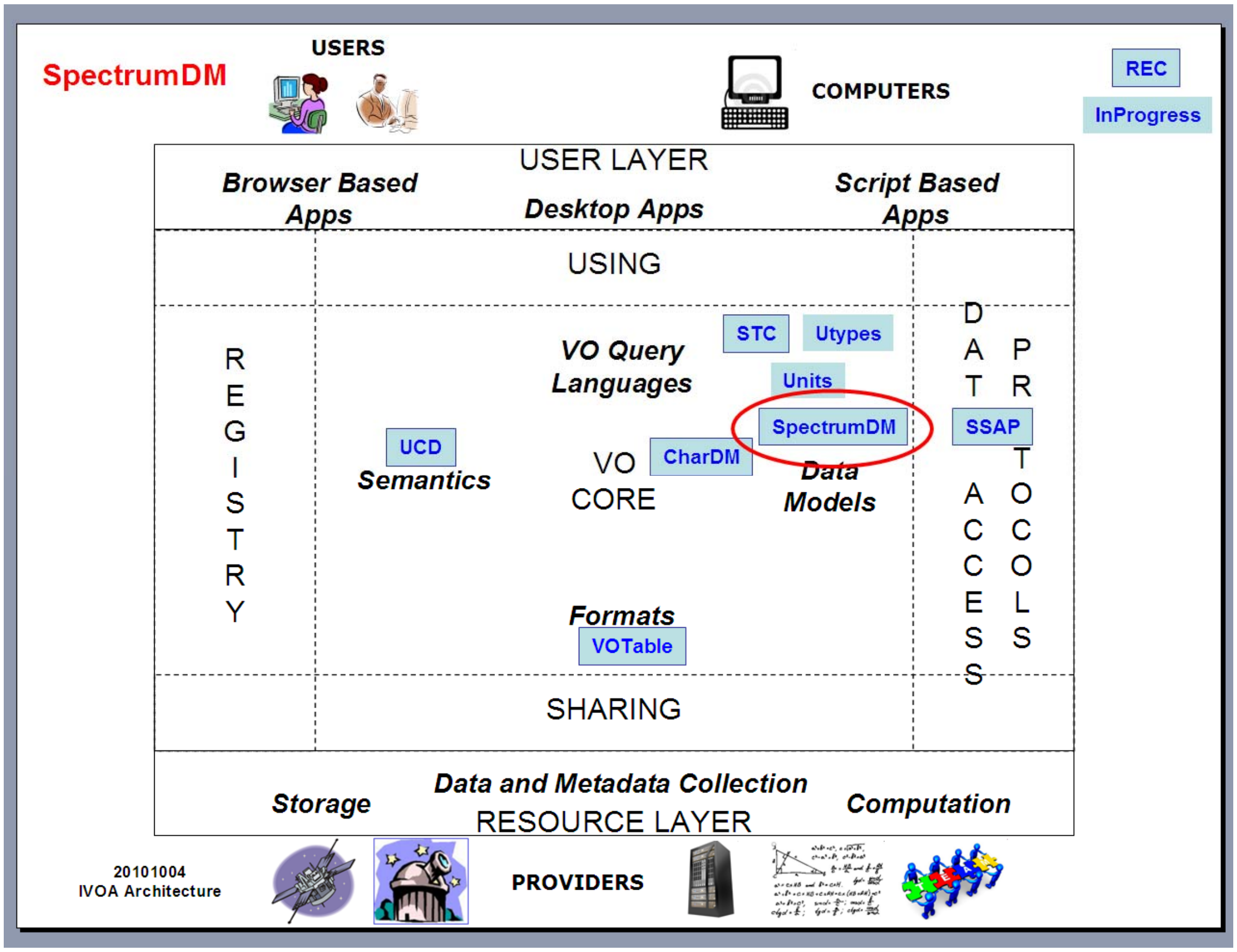

Data Models in the VO aim to define the common elements of astronomical data and metadata collections and to provide a framework for describing their relationships so these become inter operable in a transparent manner.

The Spectrum Data Model (SpectrumDM) standard presents a data model describing the structure of spectrophotometric datasets with spectral and temporal coordinates and associated metadata. This data model may be used to represent spectra, time series data, segments of SED (Spectral Energy Distributions) and other spectral or temporal associations. SpectrumDM is used with the associated Data Access Protocol, SSAP (Simple Spectra Access Protocol).

As with most of the VO Data Models, SpectrumDM makes use of STC, Utypes, Units and UCDs. Furthermore, SpectrumDM makes reference to the CharDM (Characterization Data Model). It can be serialized with a VOTable.

The SpectrumDM v1.03 became an IVOA Recommended standard in October 2007.

## 6.5 SSLDM – Simple Spectral Line DM

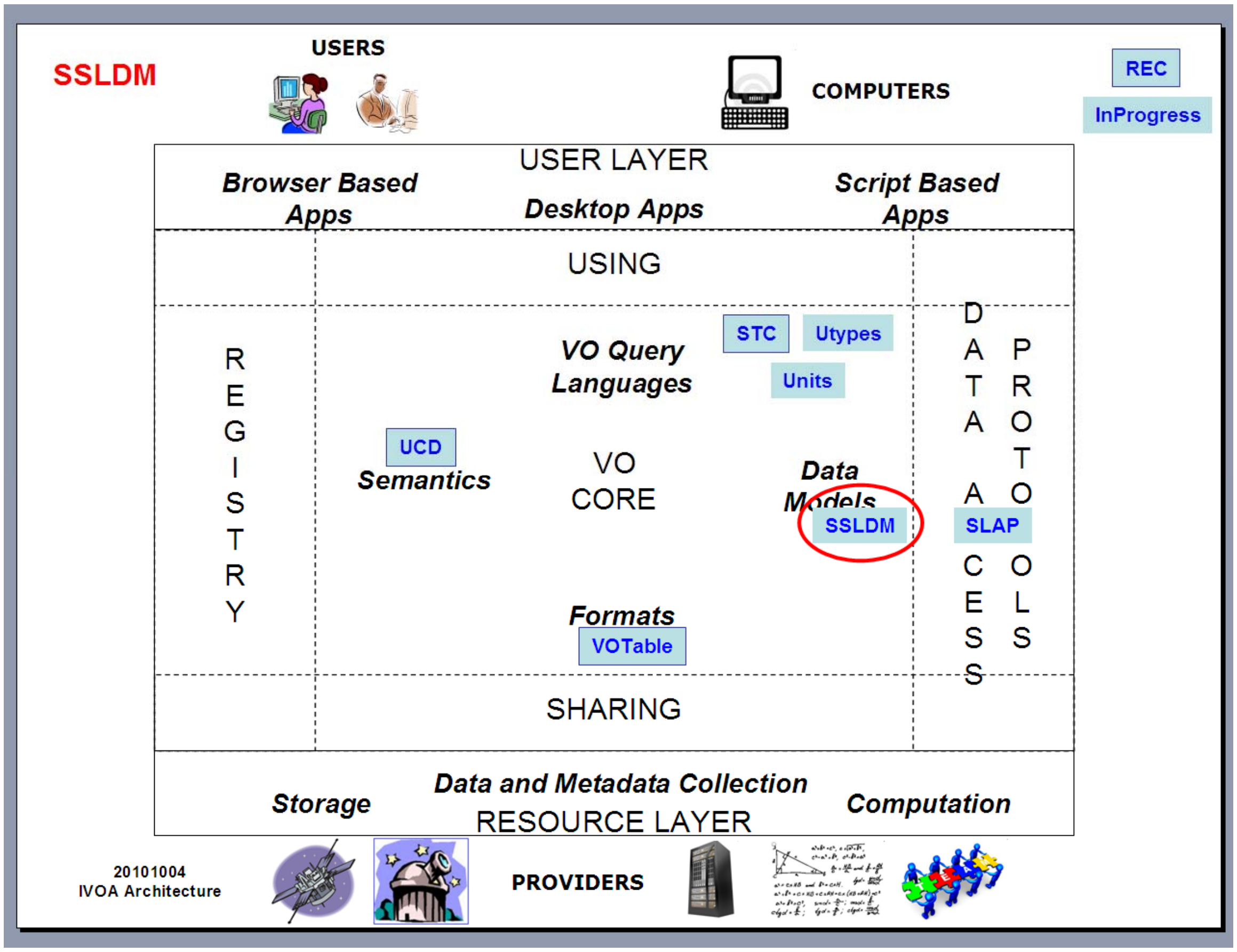

Data Models in the VO aim to define the common elements of astronomical data and metadata collections and to provide a framework for describing their relationships so these become inter operable in a transparent manner.

The Simple Spectral Line Data Model (SSLDM) standard presents a data model describing Spectral Line Transitions. In the astrophysical sense, a ***line*** is considered as the result of a ***transition*** between two ***levels***. Under the basis of this assumption, a whole set of objects and attributes have been derived to define properly the necessary information to describe lines appearing in astrophysical contexts. The SSLDM is used with the Simple Line Access Protocol (SLAP).

The SSLDM does not provide a complete description of Atomic and Molecular Physics, but makes reference to it, pending its completion.

As with most of the VO Data Models, SSLDM makes use of STC, Utypes, Units and UCDs. SSLDM can be serialized with a VOTable

SSLDM v1.0 is currently under review and it expected to become an IVOA Recommended standard in 2010.

## 6.6 CharDM – Characterization DM

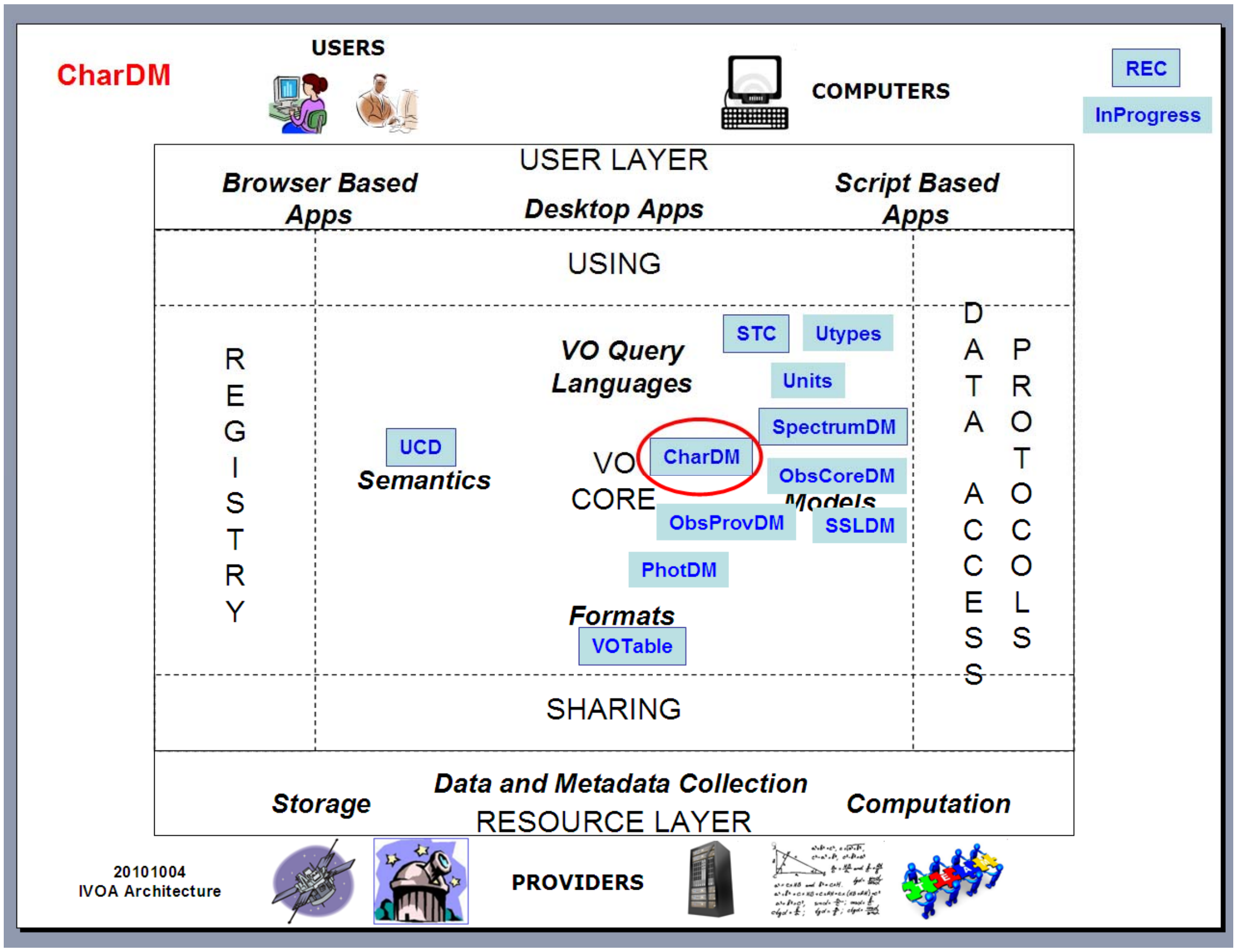

Data Models in the VO aim to define the common elements of astronomical data and metadata collections and to provide a framework for describing their relationships so these become inter operable in a transparent manner.

The Astronomical Dataset Characterization Data Model (CharDM) defines and organizes all the metadata necessary to describe how a dataset occupies multidimensional physical space, quantitatively and, where relevant, qualitatively. The model focuses on the axes used to delineate this space, including but not limited to Spatial (2D), Spectral and Temporal axes, as well as an axis for the Observable (e.g. flux, number of photons, etc.), or any other physical axes. It should contain, but is not limited to, all relevant metadata generally conveyed by FITS keywords. The Characterization Data Model is an abstraction which can be used to derive a structured description of data and thus facilitate its discovery and scientific interpretation.

Various other VO Data Models are making reference to the CharDM, in particular, ObsCoreDM (Observation Core DM), ObsProvDM (Observation and Provenance DM), SpectrumDM, SSLDM (Simple Spectral Line DM), PhotDM (Photometry DM).

As with most of the VO Data Models, CharDM makes use of STC, Utypes, Units and UCDs. CharDM can be serialized with a VOTable

CharDM v1.13 became an IVOA Recommended standard in March 2007.

## 6.7 ObsCoreDM – Observation Core DM

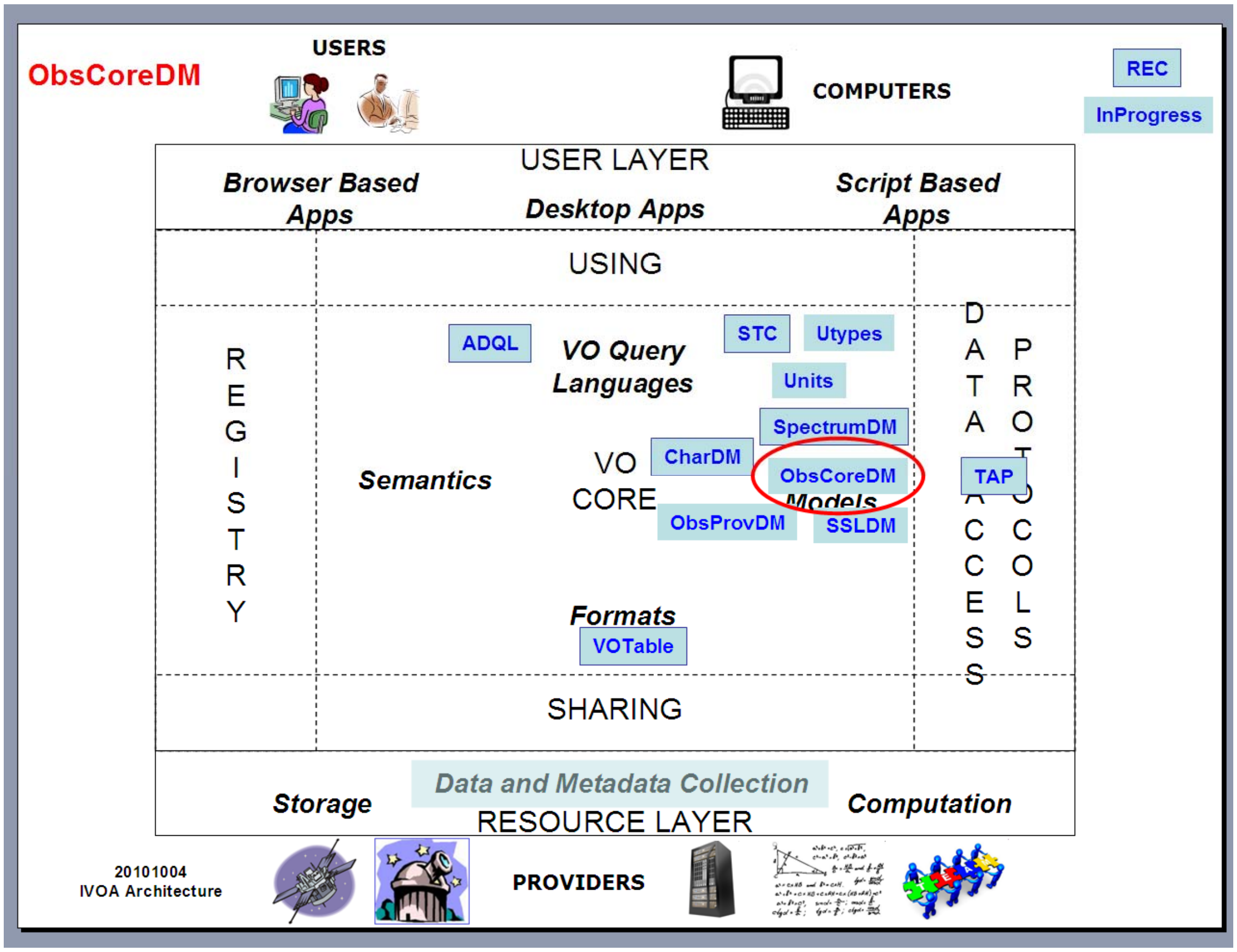

Data Models in the VO aim to define the common elements of astronomical data and metadata collections and to provide a framework for describing their relationships so these become inter operable in a transparent manner.

The Observation Core Data Model (ObsCoreDM) is meant to define the core components of all queryable metadata that play a role in the discovery of observations. It is meant to be implemented using the Utypes derived from this model in a TAP/ADQL implementation at various archives sites. It focuses on reasonable set of common readily-available metadata descriptors and thus allows for reasonable cost implementation for data providers. This has often been referred as the *ObsTAP* project.

As with most of the VO Data Models, ObsCoreDM makes use of STC, Utypes, Units and UCDs. ObsCoreDM can be serialized with a VOTable.
ObsCoreDM can make reference to more complete DM, such as ObsProvDM, CharDM, Spectrum DM or SSLDM.

The ObsCoreDM standard is still under development and is expected to become an IVOA Recommended standard in 2010-early 2011.

## 6.8 ObsProvDM – Observation and Provenance DM

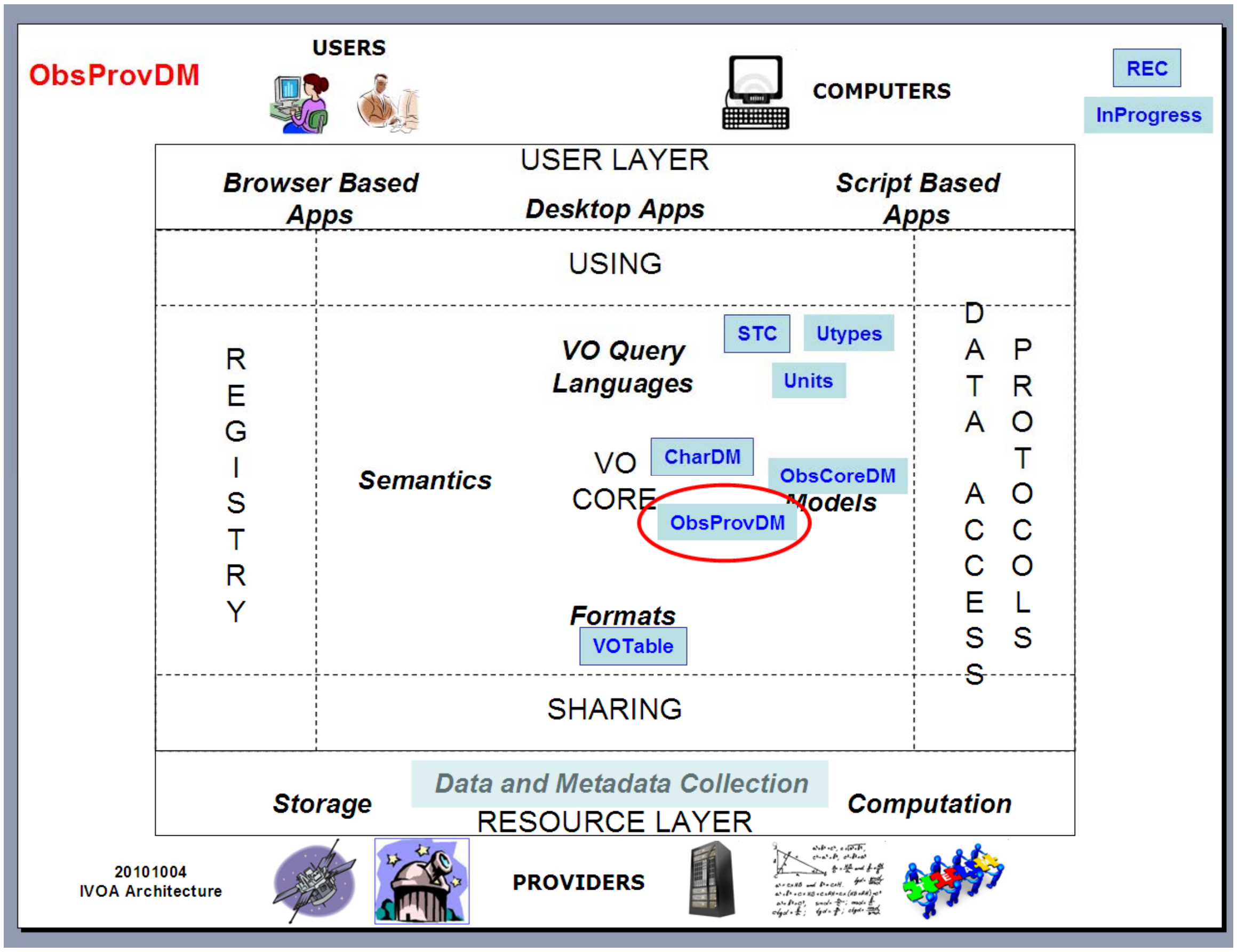

Data Models in the VO aim to define the common elements of astronomical data and metadata collections and to provide a framework for describing their relationships so these become inter operable in a transparent manner.

The Observation and Provenance Data Model (ObsProvDM) is definition of all the components of an observation, including information about its provenance, how it has been processed, under which condition the observation has been carried out, etc.

Detailed definition of ObsProvDM and its potential interdependencies with other IVOA standards still need to be defined. It is an extension of the simple ObsCoreDM and it will most likely make reference to the CharDM.

As with most of the VO Data Models, ObsProvDM will make use of STC, Utypes, Units and UCDs. ObsProvDM will be serializable with a VOTable

In 2010, the ObsProvDM standard is still under development.

## 6.9 PhotDM – Photometry DM

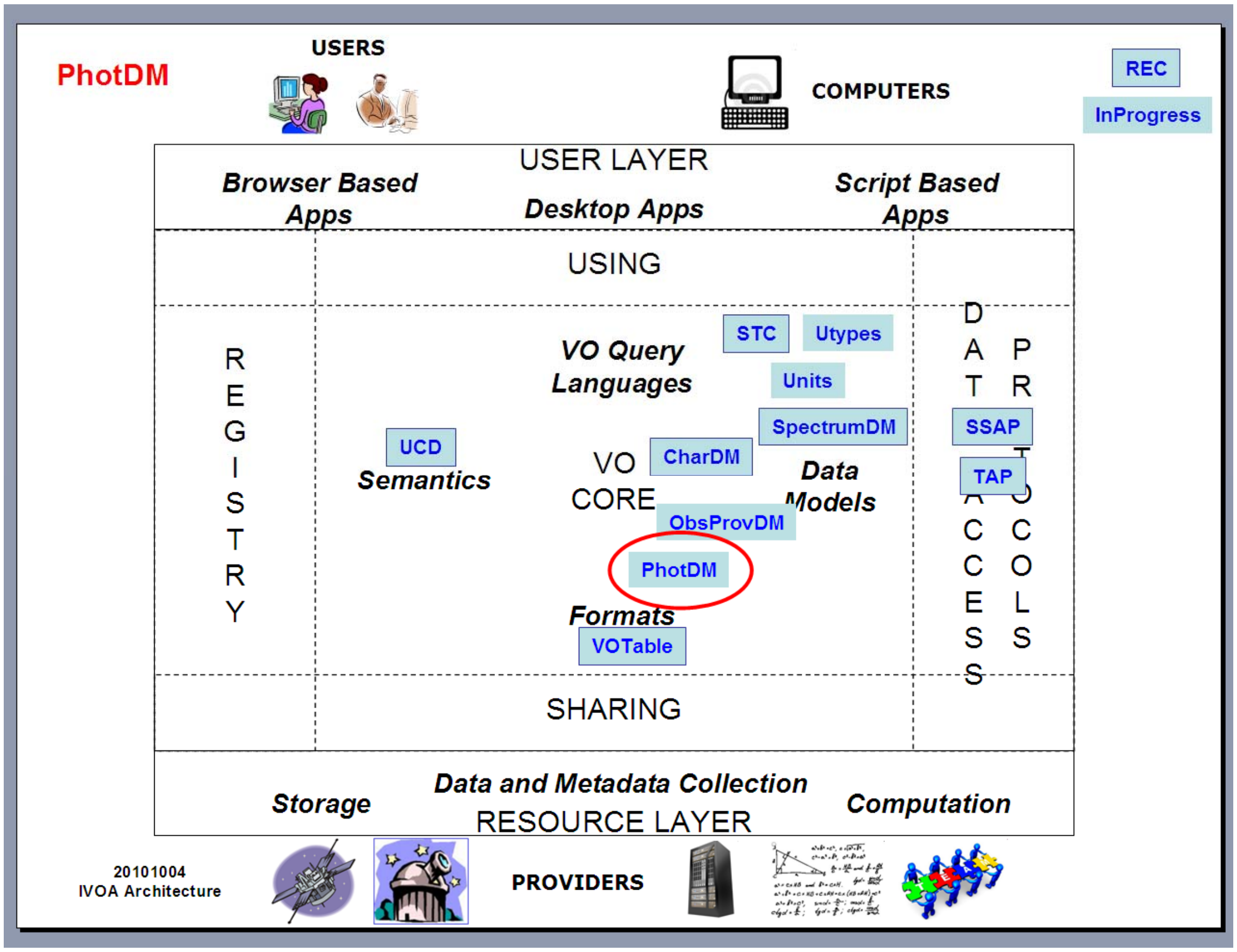

Data Models in the VO aim to define the common elements of astronomical data and metadata collections and to provide a framework for describing their relationships so these become inter operable in a transparent manner. Photometry is a technique of astronomy concerned with measuring the flux, or intensity of an astronomical object's electromagnetic radiation.

The Photometry Data Model (PhotDM) standard describes photometric filters through a simple data model in order to allow the creation of protocols to access photometric data in magnitudes. PhotDM could be used in conjunction with other IVOA Data Access Protocol such as SSAP (Simple Spectra Access Protocol) or TAP (Table Access Protocol).

PhotDM makes reference to the CharDM (Characterization DM), to the SpectrumDM and to the provenance part of the ObsProvDM (Observation and Provenance DM).

As with most of the VO Data Models, PhotDM will make use of STC, Utypes, Units and UCDs. PhotDM will be serializable with a VOTable.

The PhotDM standard is still under development and is expected to become an IVOA Recommended standard in 2010-early 2011.

## 6.10 SimDM – Simulations DM

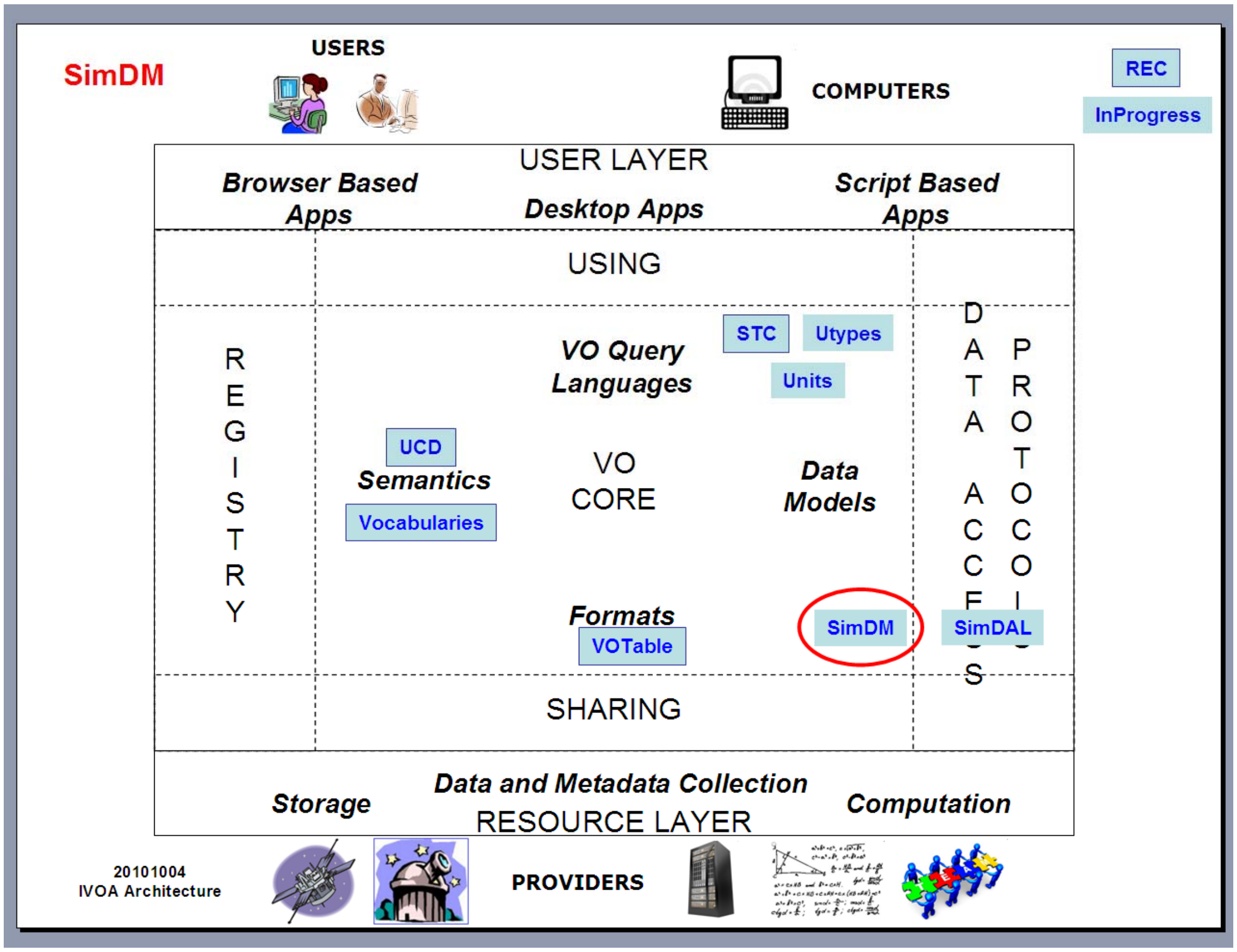

Data Models in the VO aim to define the common elements of astronomical data and metadata collections and to provide a framework for describing their relationships so these become inter operable in a transparent manner.
The Simulation Data Model (SimDM) defines and organizes all the metadata necessary to describe simulation datasets. SimDM can be used in conjunction with SimDAL to enable uniform access to simulation and theoretical products.

In 2010, the SimDM standard is still under development.. Therefore, detailed definition of SimDM and its potential interdependencies with other IVOA standards still need to be defined, so the above diagram is based on relationship from other VO Data Models.

As with most of the VO Data Models, SimDM is expected to make use of STC, Utypes, Units and UCDs. SimDM will be serializable with a VOTable. SimDM may also need to develop its own Vocabulary.

## 6.11 VOEvent

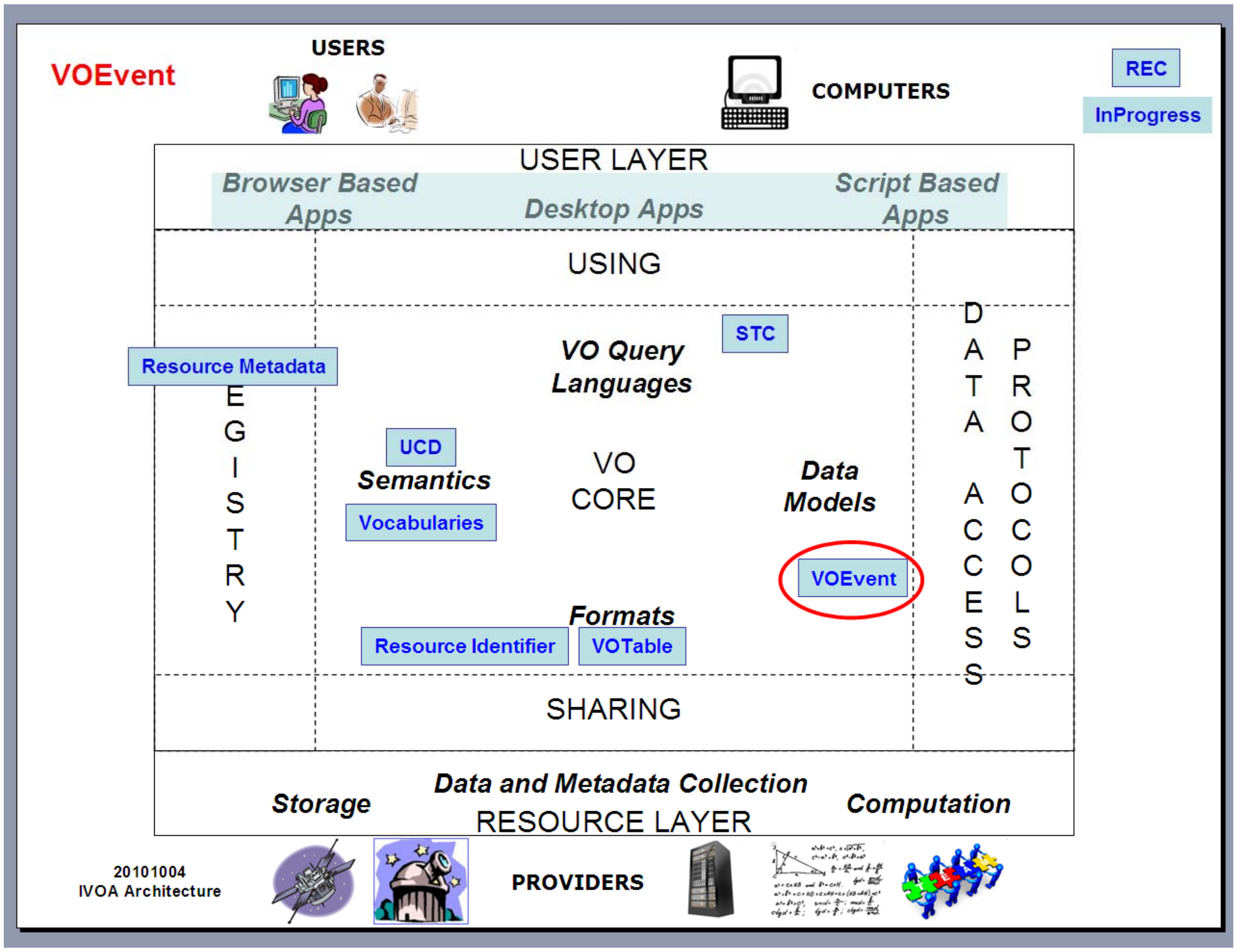

Data Models in the VO aim to define the common elements of astronomical data and metadata collections and to provide a framework for describing their relationships so these become inter operable in a transparent manner.

VOEvent defines the content and meaning of a standard information packet for representing, transmitting, publishing and archiving the discovery of a transient celestial event, with the implication that timely follow-up is being requested. The objective is to motivate the observation of targets-of-opportunity, to drive robotic telescopes, to trigger archive searches, and to alert the community. VOEvent is focused on the reporting of photon events, but events mediated by disparate phenomena such as neutrinos, gravitational waves, and solar or atmospheric particle bursts may also be reported.

VOEvent is a pragmatic effort that crosses the boundary between the Virtual Observatory and the larger astronomical community. The results of astronomical observations using real telescopes must be expressed using the IVOA VOEvent standard, be recorded and transmitted via registries and aggregators within and outside the VO, and then be captured and filtered by subscribing VO clients.

Each event that survives rigorous filtering can then be passed to other real (or possibly virtual) telescopes, for instance via RTML, to acquire real follow-up observations that will confirm (or deny) the original hypothesis as to the classification of the object(s) or processes that generated that particular VOEvent in the first place. This must happen quickly (often within seconds of the original VOEvent) and must minimize unnecessary expenditures of either real or virtual resources.

VOEvent is a mechanism for broadcasting discoveries that others may wish to follow-up, and this purpose defines its scope. An astronomical discovery that cannot benefit from immediate follow-up is not a good candidate for expression as a VOEvent.

VOEvent servers and streams will be registered through a Resource Identifier and will be compatible with IVOA Resource Metadata.

VOEvent packets work in concert with external content such as VOTables, and rely on IVOA features such as UCDs and Vocabularies. Spatial and temporal coordinate systems and errors are to be described with STC.

VOEvent v1.11 became an IVOA Recommended standard in November 2006.

# 7 VO Core: Query Language Standards

## 7.1 ADQL – Astronomical Data Query Language

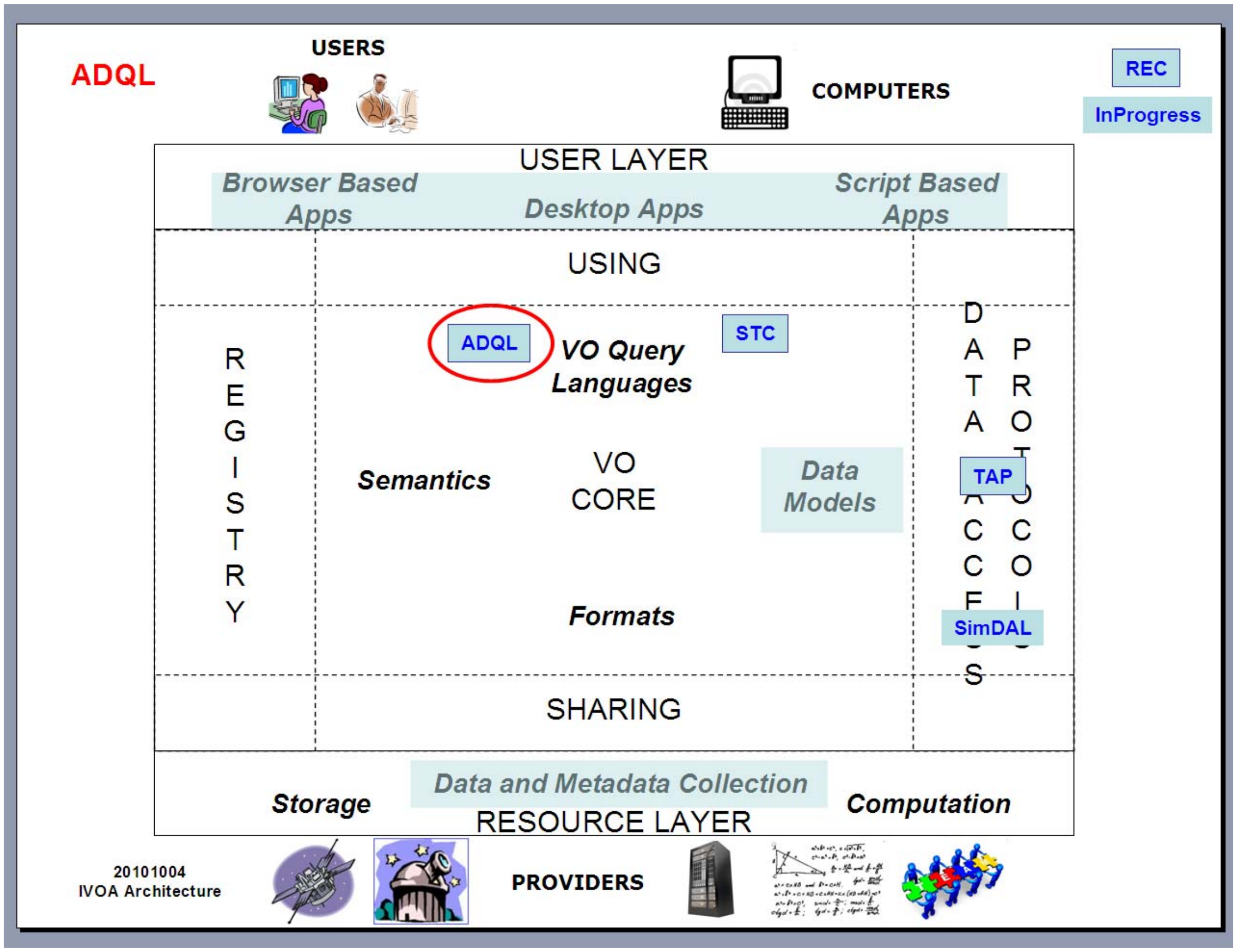

Simple Data Access Protocols (ie SIAP, SCS) have defined single table query, but there is a need for more powerful query languages in the VO.

The Astronomical Data Query Language (ADQL) is such a language. ADQL is based on the Structured Query Language (SQL), specifically SQL 92. The VO has a number of tabular data sets and many of them are stored in relational databases, making SQL a convenient means of access. A subset of the SQL grammar has been extended to support queries that are specific to astronomy. In practice the geometry functions of ADQL are defined by the IVOA STC standard.

ADQL may be required or optional in newer data access protocols standards (eg TAP, SimDAL) make direct reference to ADQL. Coupled with data models, VO Applications can build ADQL queries to access VO resources (data and metadata collections) in a more powerful and expressive ways.

The ADQL standard v2.0 became an IVOA Recommended Standard in October 2008.

(Note that the Registry Interface standard makes use of an earlier version of ADQL (v1.01) which had never become an IVOA Recommendation. Hence, the Registry Interface includes an Annex containing the definition of this “ADQL 1.01” which does not correspond to the formal Recommended ADQL v2.0.)

## 7.2 PQL – Parameterized Query Language

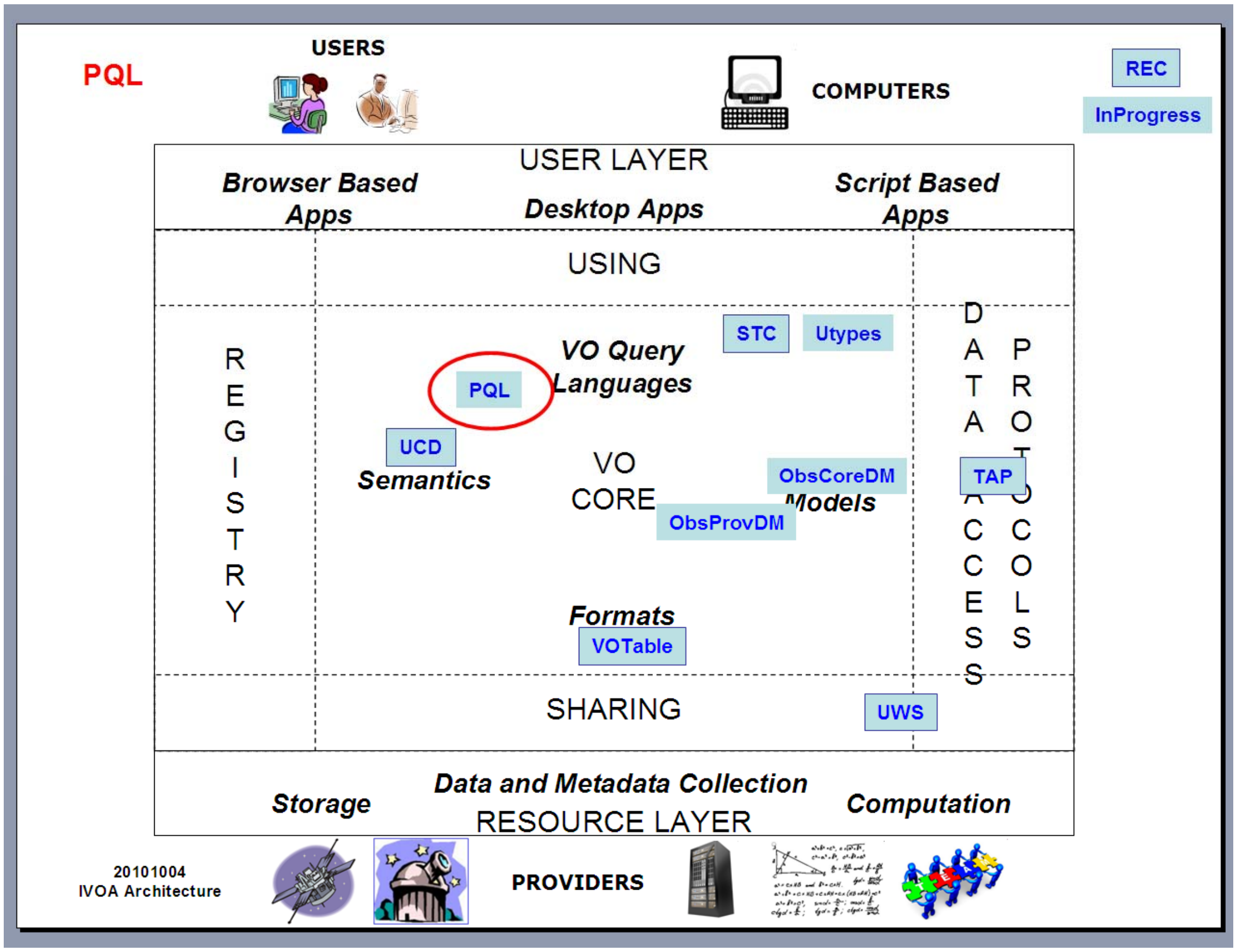

Simple Data Access Protocols (ie SIAP, SCS) have defined single table query, but there is a need for more powerful query languages in the VO.

The Parameterized Query Language (PQL) is a language used to represent simple astronomy queries posted to VO services. PQL has been developed for simple parameter-based table queries as part of the IVOA Table Access Protocol (TAP). PQL specification formalizes the syntax and meaning of PQL as a general parameter-based query language for querying tabular data.

PQL is based on past use and parameter-based query interfaces in DAL services such as Simple Image Access (SIA), Simple Spectral Access (SSA), and Simple Cone Search (SCS). Parametric queries are simple to express and to implement for cases where the data model is sufficiently well defined and adequate for the data to be queried, hiding many of the details required to pose and evaluate the query. Potential PQL relationship with IVOA Data Models, such as ObsCoreDM and ObsProvDM still has to be clarified.

Query parameters in PQL are defined through Utypes or UCDs, making also use of STC. Output of a TAP/PQL query is typically in VOTable.

By default parameter queries execute synchronously, and upon successful execution return the output table directly to the client. Parameter queries may also execute asynchronously using the UWS mechanism provided by the main TAP service.

In 2010, the PQL standard is still under development.

# 8 VO Core: Semantics

## 8.1 UCD1+ – Unified Content Descriptor

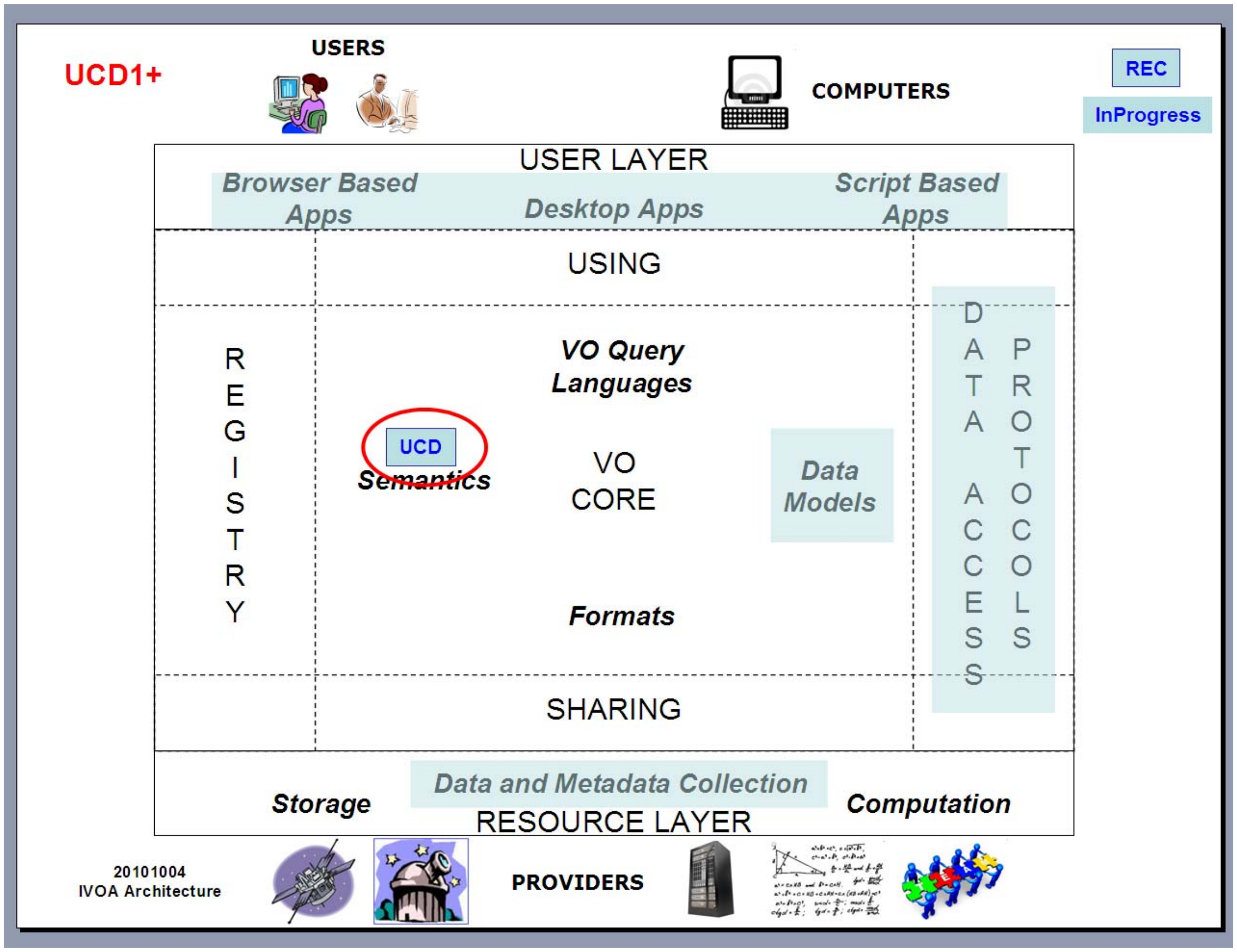

The UCD1+ (Unified Content Descriptor) is part of the VO Core standards. The UCD1+ standard presents the IVOA controlled vocabulary for describing astronomical data quantities. UCD 1+ represents an evolution to the initial UCD1. The UCD1+ standard document also addresses the questions of maintenance and evolution of the UCD1+ controlled vocabulary. It allows a common vocabulary to represent data and metadata collections, facilitating metadata discovery and analysis from VO Applications.

As a central part of the VO core, UCD1+ is being used in most of the other VO standards (ie Data Models, Data Access Protocols).

UCD v1.1 has initially become an IVOA Recommended standard in August.
UCD 1+ v1.13 has then superseded the previous recommendation in April 2007.

## 8.2 Vocabularies

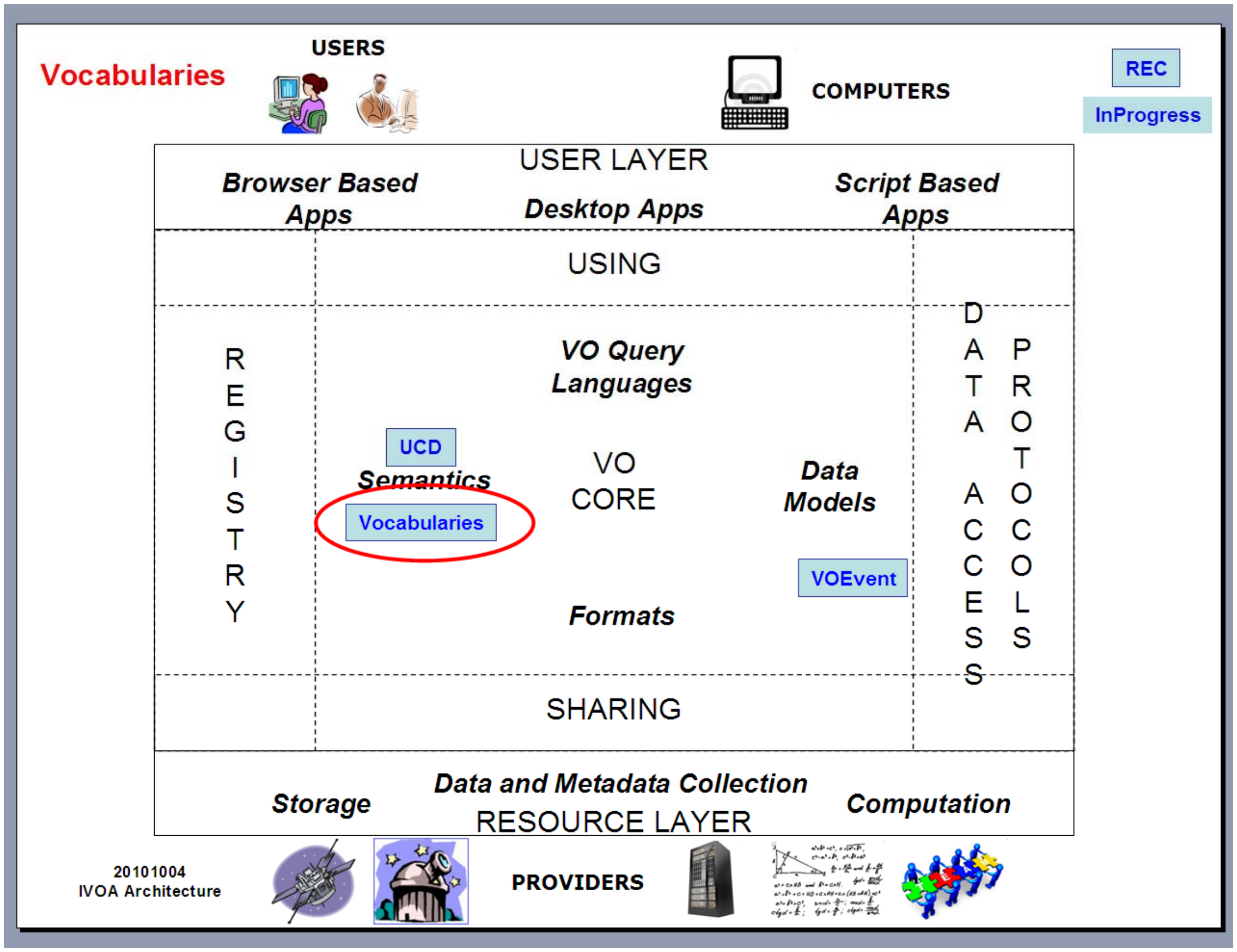

Vocabularies is part of the VO Core standards that could be used in various VO areas.

Astronomical information of relevance to the VO is not confined to quantities easily expressed in a catalogue or a table. Fairly simple things such as position on the sky, brightness in some units, times measured in some frame, redshifts, classifications or other similar quantities are easily manipulated and stored in VOTable and can currently be identified using IVOA UCDs. However, astrophysical concepts and quantities use a wide variety of names, identifications, classifications and associations, most of which cannot be described or labeled via UCDs.

This Vocabularies standard specifies a standard format for the VO to define it own vocabulary. This standard is based on the W3C's Resource Description Framework (RDF) and Simple Knowledge Organization System (SKOS). By adopting a standard and simple format, the IVOA will permit different groups to create and maintain their own specialized vocabularies while letting the rest of the astronomical community access, use, and combine them. The use of current, open standards ensures that VO applications will be able to tap into resources of the growing semantic web.

Example of VO vocabularies that could be defined are UCD vocabulary, VOEvent, Constellation name, etc…

Vocabularies 1.19 became an IVOA Recommended Standard in October 2009.

# 9 VO Core: Formats Standards

## 9.1 VOTable

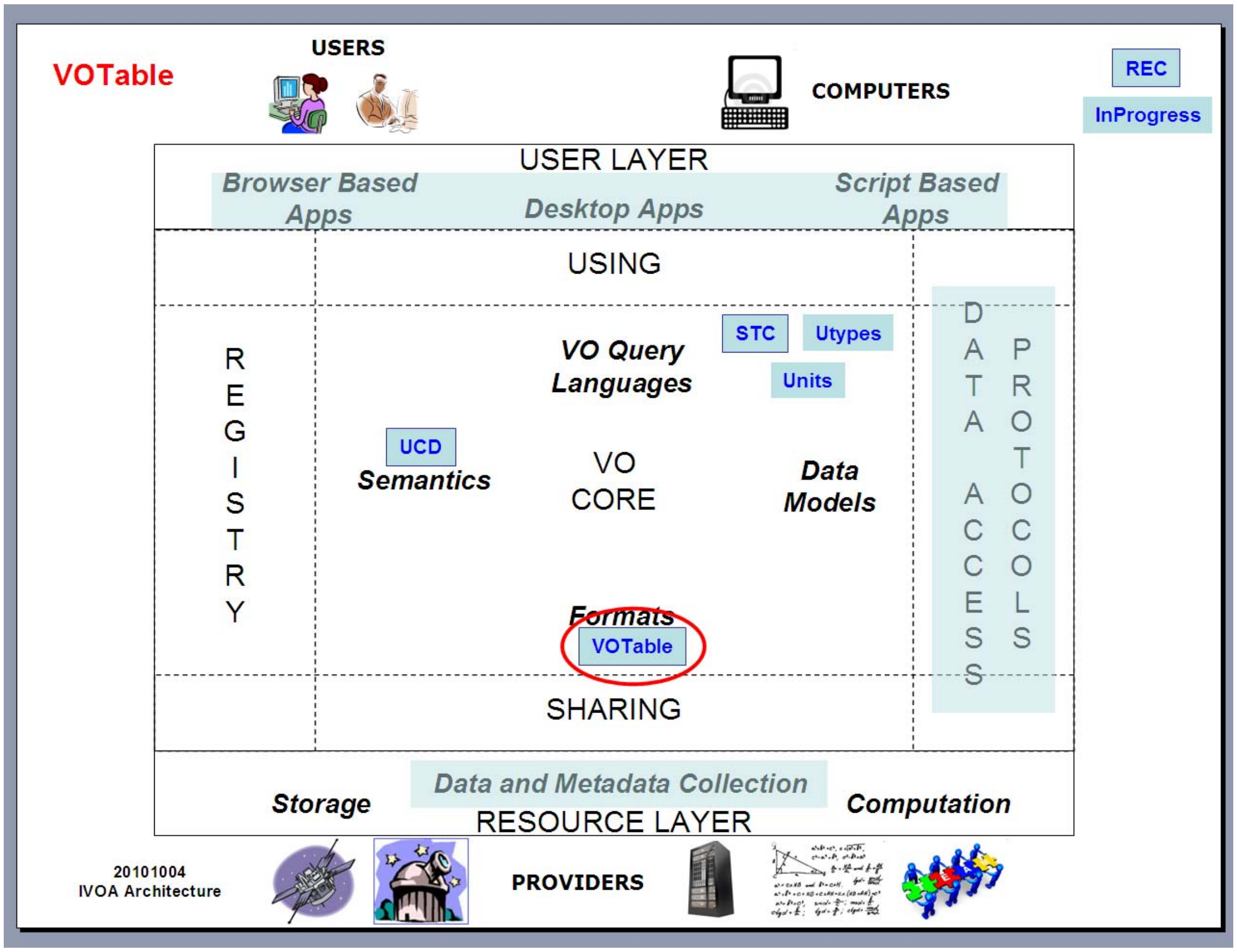

The VOTable standard is part of the VO core standards that are used in various other VO areas. VOTable format has been designed as a flexible storage and exchange format for tabular data, with particular emphasis on astronomical tables. Based on the XML standard, it also has been designed to remain close to the widely used astronomical standard FITS binary table format.

VOTable has been the very first IVOA Recommended standard, back in August 2004 as version 1.1. Later on, it has been upgraded to v1.2 in November 2009.

VOTable allows metadata and data to be stored separately. As such, VOTable is widely used in various VO contexts, in particular is can be used to store and exchange data and metadata collection, in all the Data Access Protocol, and as well in the VO applications.
VOTable makes use of UCD, to describe the names of the table column, as well at STC to make reference to Space Time and Coordinate systems.

When they become an IVOA recommended standard, VOTable will be able to make more systematic use of Units and Utypes.

## 9.2 Resource Identifier

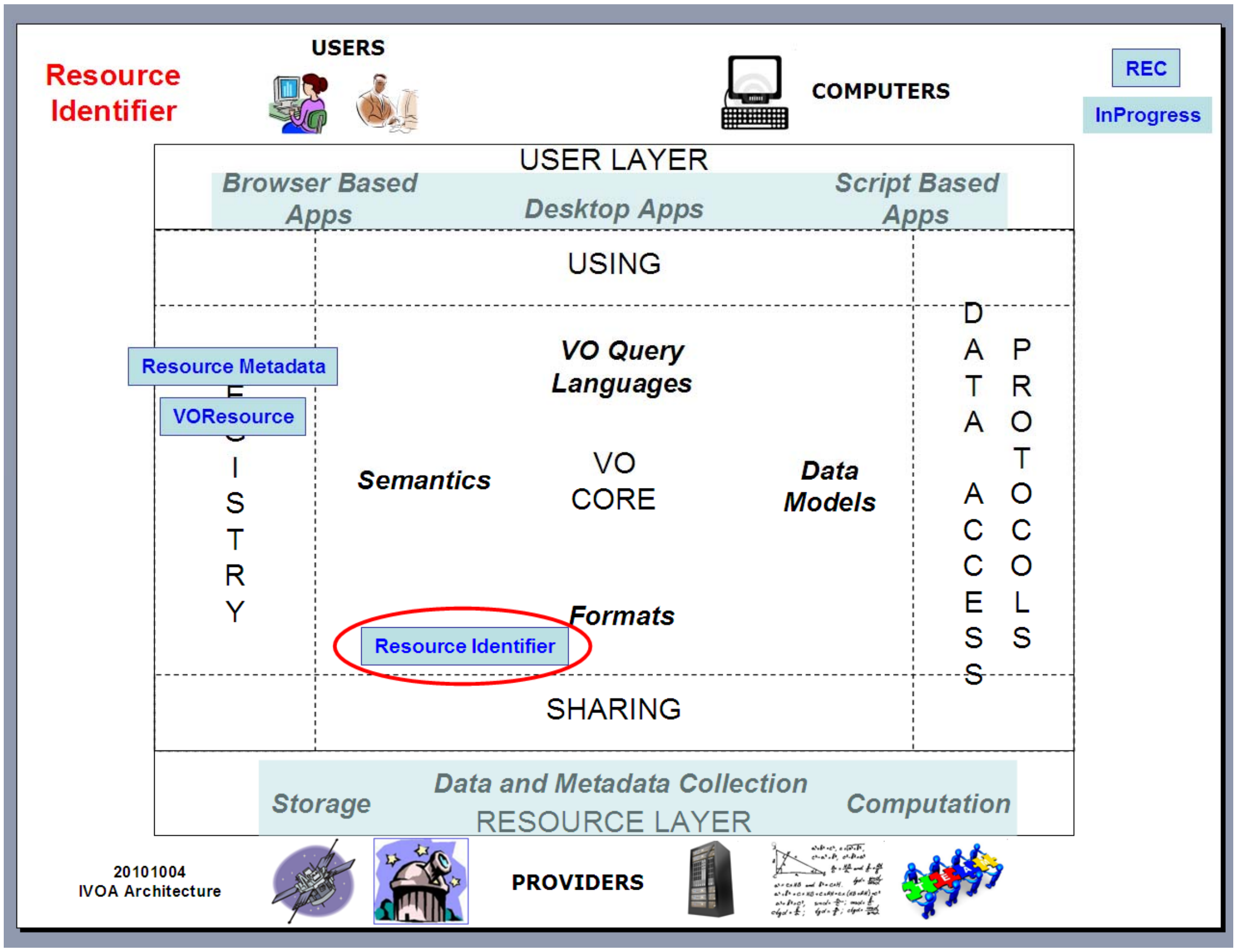

The Resource Identifier is part of the VO Core standards. An IVOA Resource Identifier is a globally unique name for a VO resource (such as a data or a metadata collection, a VO Storage element, or again a VO Computation service). This name can be used to retrieve a unique description of the resource from an IVOA-compliant registry. Identifiers are very important to registries as they aid users in discovering data and services. In general, a registry stores descriptions of data and services in a searchable form, and it distinguishes them by a unique ID. Furthermore, when users encounter an ID, they should be able to go to a registry and find out something about the thing it refers to.
The Resource Identifier can be described with the IVOA Resource Metadata.

The Resource Identifier v1.12 became an IVOA Recommended standard in March 2007.

# 10 Using: Applications Related Standards

## *10.1 SAMP – Simple Application Messaging Protocol*

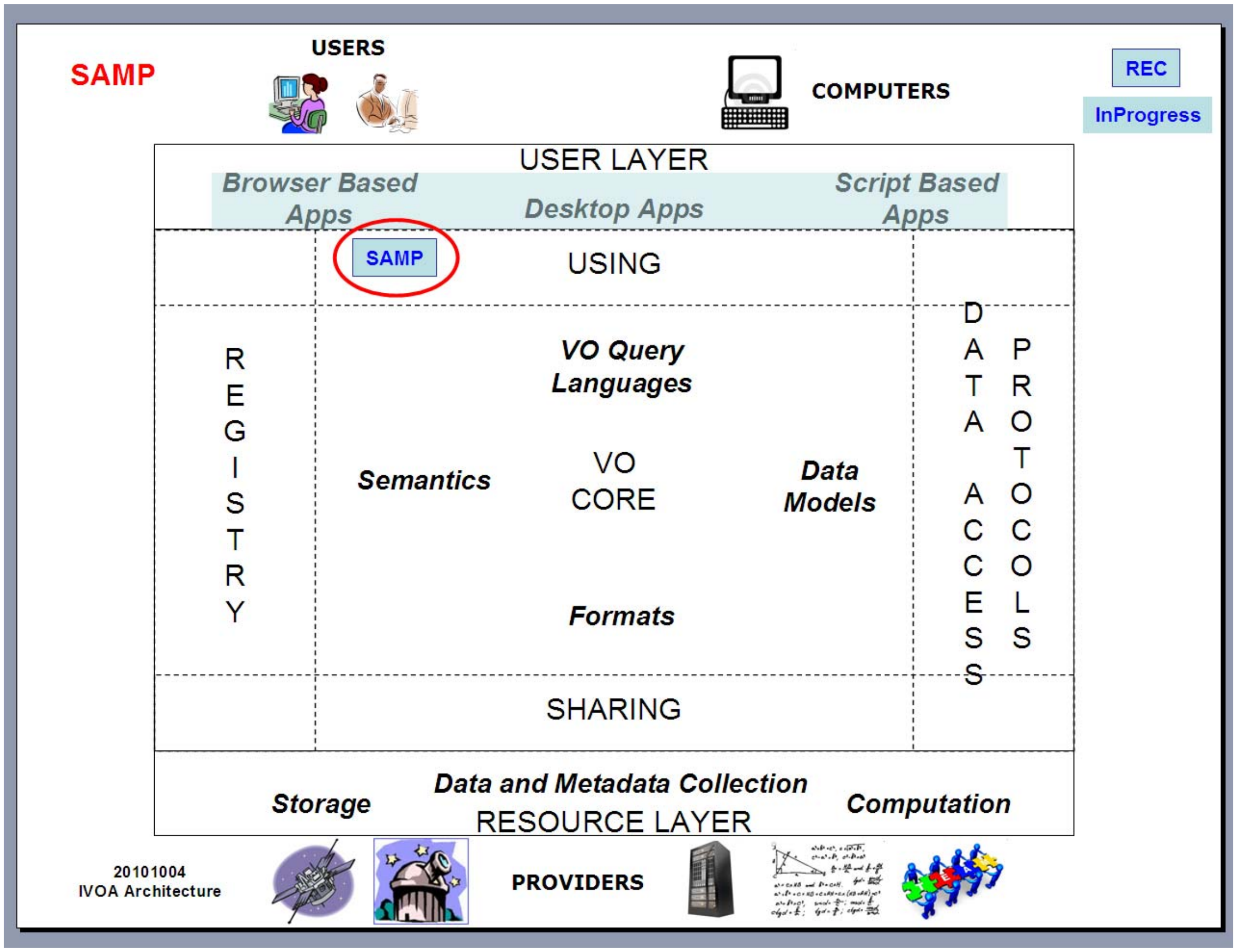

VO partners have recognized that building a monolithic tool that attempts to fulfill all the requirements of all users is impractical, and it is a better use of resources to enable individual tools to work together better. One element of this is defining common file formats for the exchange of data between different applications. Another important component is a messaging system that enables the applications to share data and take advantage of each other's functionality.
SAMP, the Simple Application Messaging Protocol, is a standard for allowing software tools to exchange control and data information, thus facilitating tool interoperability, and so allowing users to treat separately developed applications as an integrated suite.

An example of an operation that SAMP might facilitate is passing a source catalogue from one GUI application to another, and subsequently allowing sources marked by the user in one of those applications to be visible as such in the other.

The protocol has been designed, and implementations developed, within the context of the International Virtual Observatory Alliance (IVOA), but the design is not specific either to the Virtual Observatory (VO) or to Astronomy. It is used in practice for both VO and non-VO work with astronomical tools, and is in principle suitable for non-astronomical purposes as well.

The SAMP standard itself is neither a dependent, nor a dependency, of other VO standards, but it provides valuable glue between user-level applications which perform different VO-related tasks, and hence contributes to the integration of Virtual Observatory functionality from a user's point of view. Most existing tools which operate in the User Layer of this architecture provide SAMP interoperability.

SAMP was built on a prior messaging protocol, called PLASTIC which had been in use since 2006, but PLASTIC has largely been superseded by SAMP.

SAMP v1.11 became an IVOA Recommended standard in April 2009.

An updated version of SAMP (v1.2) should become an IVOA Recommended standard in 2010.

## 10.2 SSO – Single Sign On

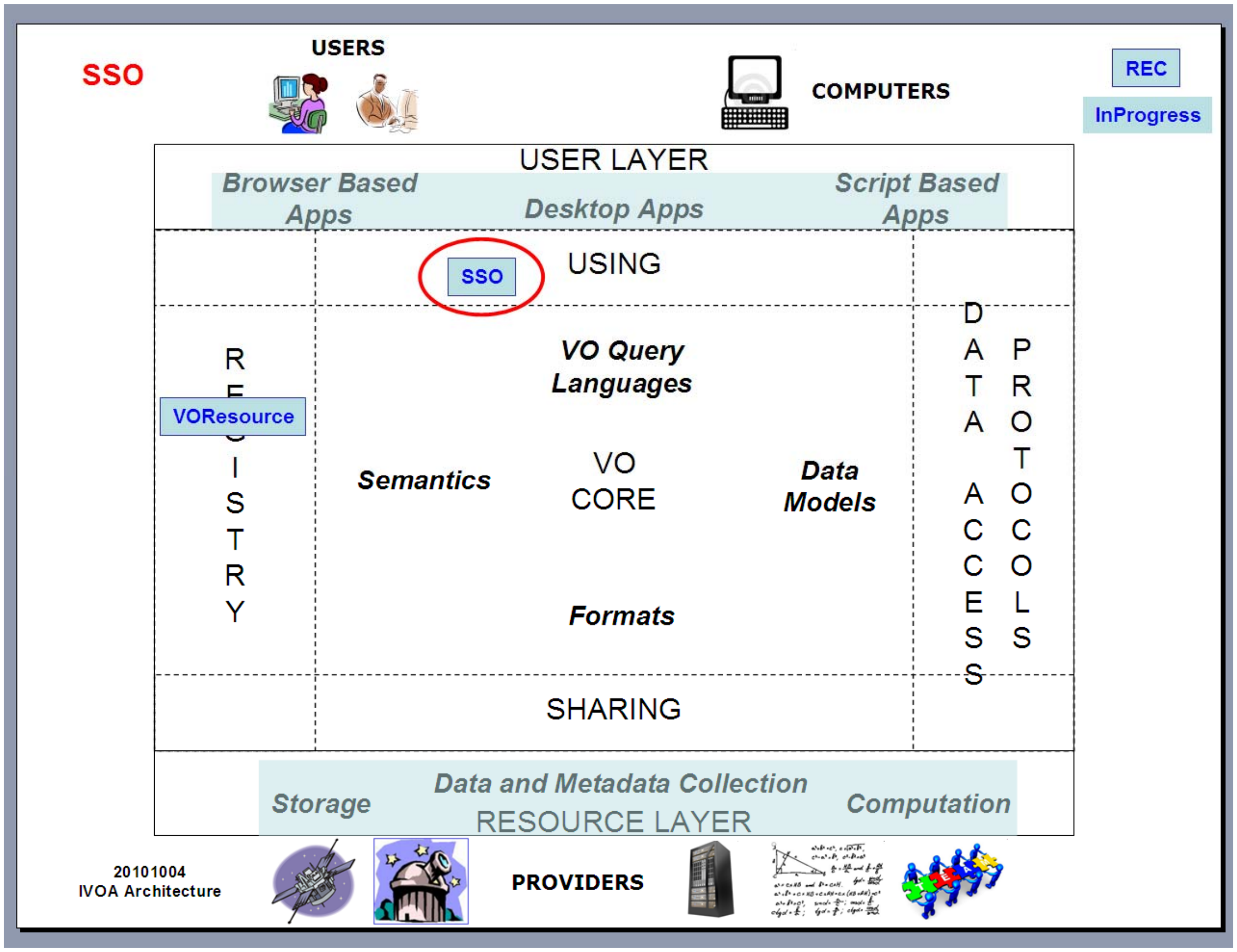

Although most of the datasets in astronomy and in the VO are of the public domain, there are some use cases that require access control such as when data are under a proprietary period or when data is to be privately shared amongst a group of collaborating researchers.

Hence some standards are required within the User Layer to enable user authentication to non-public datasets and storage elements and to enable interoperability amongst VO applications (“Using”).

The IVOA Single Sign On (SSO) standard describes the user authentication mechanisms (existing industry standards) to access VO Resources. IVOA SSO architecture is a system in which users assign cryptographic credentials to user agents so that the agents may act with the user’s identity and access rights. This standard describes how agents use those credentials to authenticate the user’s identity in requests to services.

When a service is registered in an IVO registry through a VOResource, that service's resource document may include metadata expressing conformance to one or more of the authentication mechanisms approved in the IVOA SSO profile. Such a service must implement those mechanisms as described in this document, and clients of the service must participate in the mechanism when calling the service.

SSO v1.01 became an IVOA Recommended standard in January 2008.

## 10.3 CDP – Credential Delegation Protocol

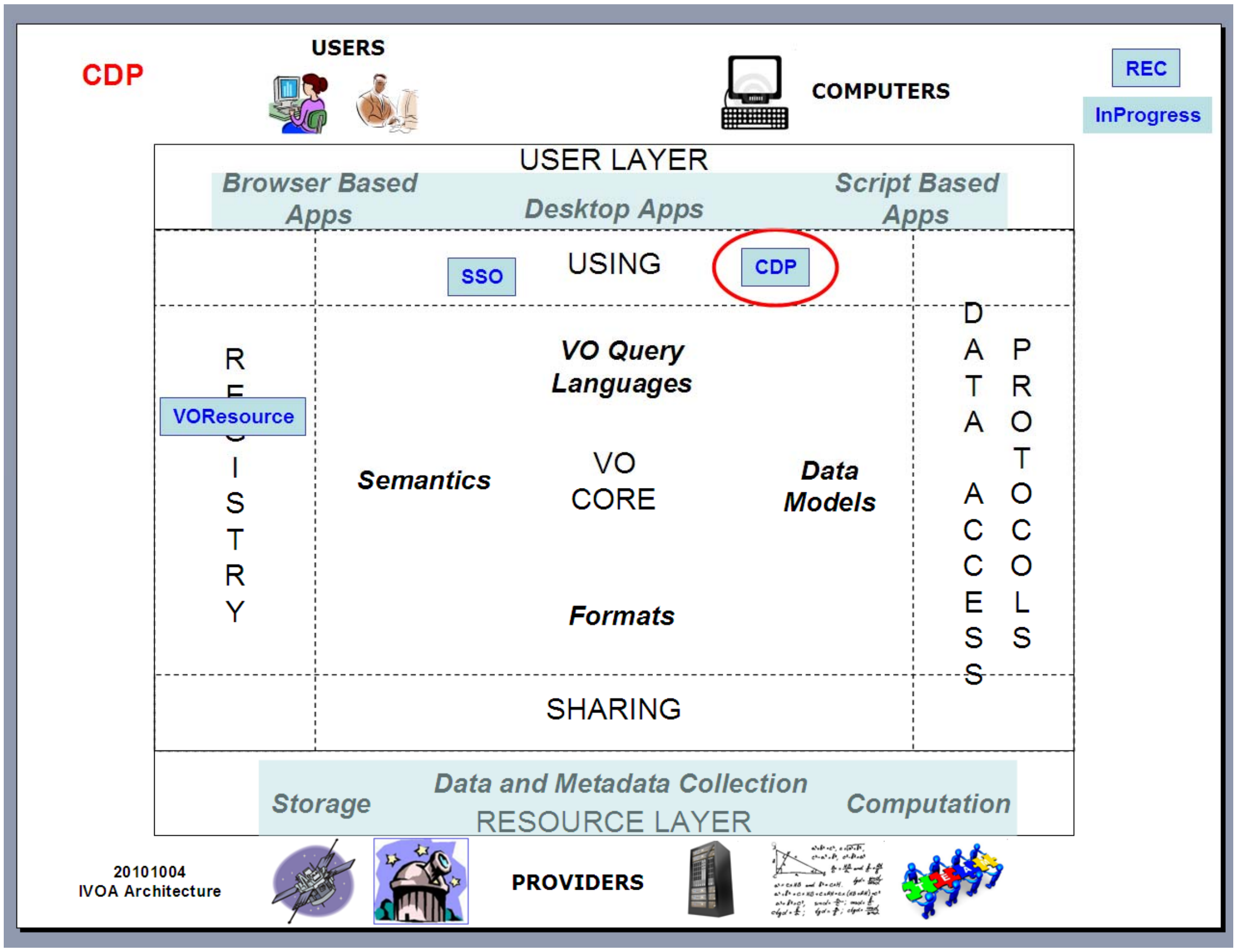

Although most of the datasets in astronomy and in the VO are of the public domain, there might be some which require access control as they are under a proprietary period. Another use case can be for a group of people who want to share datasets amongst themselves only.
Hence some standards are required by the User Layer to enable user authentication to proprietary datasets and storage elements as well as interoperability amongst VO applications (“Using”).

The IVOA Credential Delegation Protocol (CDP) allows a client program to delegate a user's credentials (usually defined in SSO) to a service such that that service may make requests of other services in the name of that user.

When a service is registered in an IVO registry through a VOResource, that service’s resource document may include metadata expressing conformance to the Credential Delegation Protocol.

CDP v1.0 became an IVOA Recommended standard in February 2010.

## 10.4 WS BP – Web Service Basic Profile

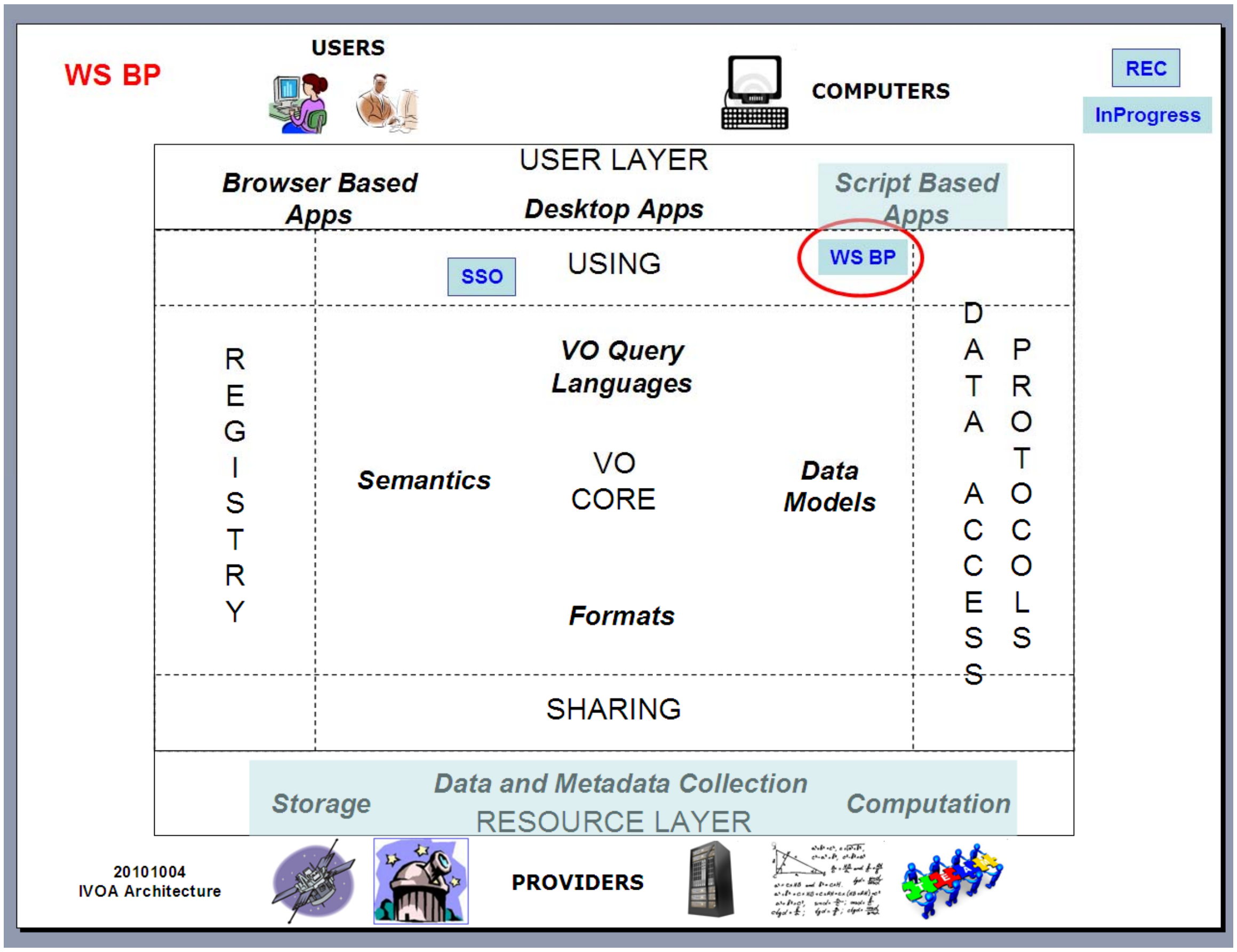

Web Services are now a common approach to providing access to Internet resources. It is foreseeable that many VO *providers* will provide VO services in this manner.

The IVOA Web Services Basic Profile (WS BP) provides to the VO web services providers a guideline on how to use the existing specifications in the IVOA web services context. IVOA WS BP is based on the specifications of the Web Services Interoperability (WS-I) organization, an open industry effort chartered to promote Web Services interoperability across platforms, applications, and programming languages.

The basic security profile is defined by the IVOA Single-Sign-On Profile

WS BP v1.0 is currently under review and should become a Recommendation in 2010.

# 11 Sharing Related Standards

## 11.1 VOSpace

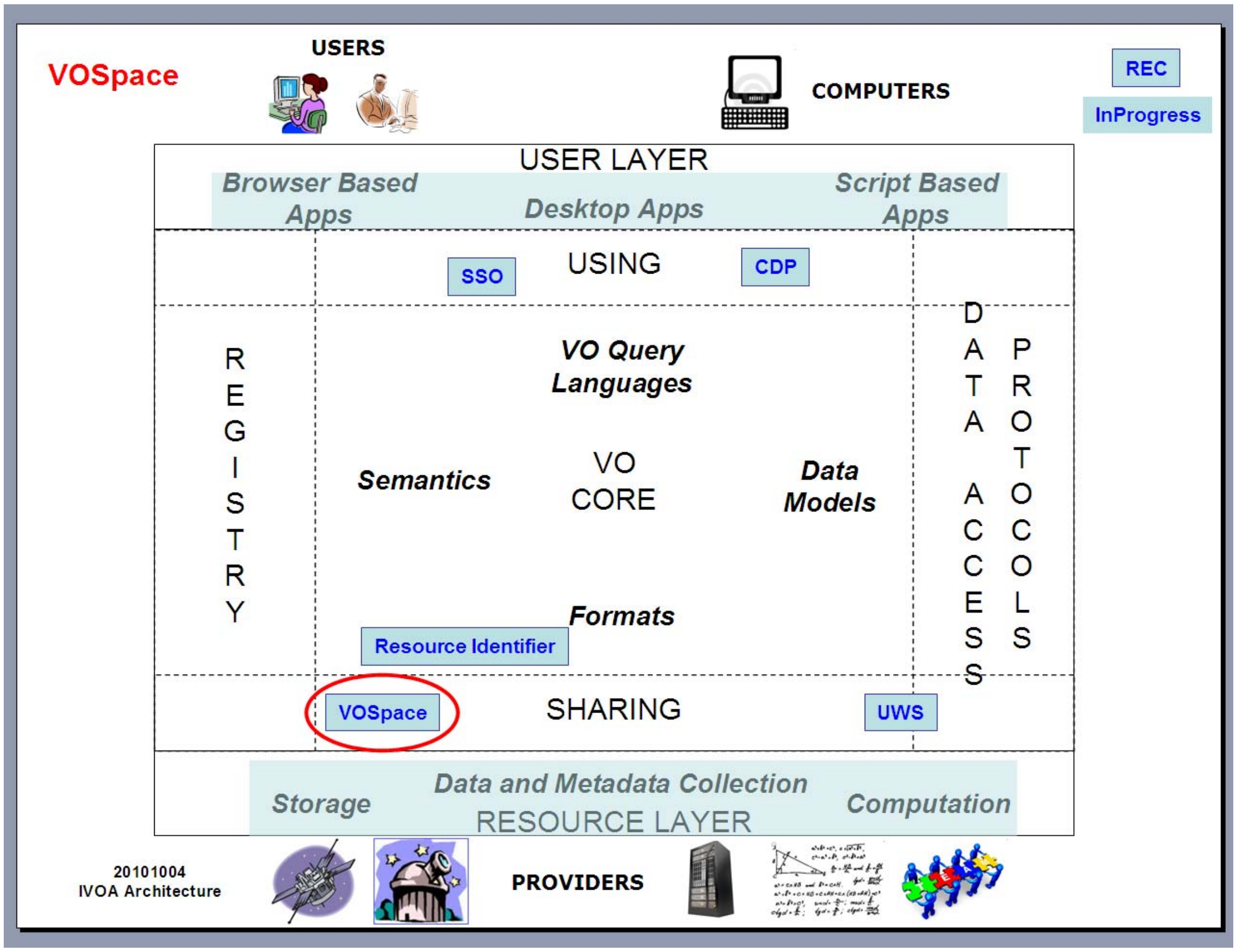

The VO provides a technical framework for the providers and the users to share their data and services ("Sharing").

VOSpace is the IVOA interface to distributed storage. It specifies how VO agents and applications can use network attached data stores to persist and exchange data in a standard way. A VOSpace web service is an access point for a distributed storage network. Through this access point, a client can:

- add or delete data objects
- manipulate metadata for the data objects
- obtain URIs through which the content of the data objects can be accessed

VOSpace does not define how the data is stored or transferred, only the control messages to gain access. Thus, the VOSpace interface can readily be added to an existing storage system.

Where the access policy requires authentication of callers, the VOSpace service shall support the IVOA authentication methods such as Single Sign On profile (SSO) or Credential Delegation Protocol (CDP).

VOSpace 1.15 currently does not make use of Universal Worker Service (UWS), but future VOSpace version will make use of it to enable elaborated job workflows involving asynchronous storage on VOSpace elements.

Once registered, a VOSpace service will get a unique IVOA Resource Identifier.

VOSpace v1.15 became an IVOA Recommended standard in October 2009.

## 11.2 VOPipe

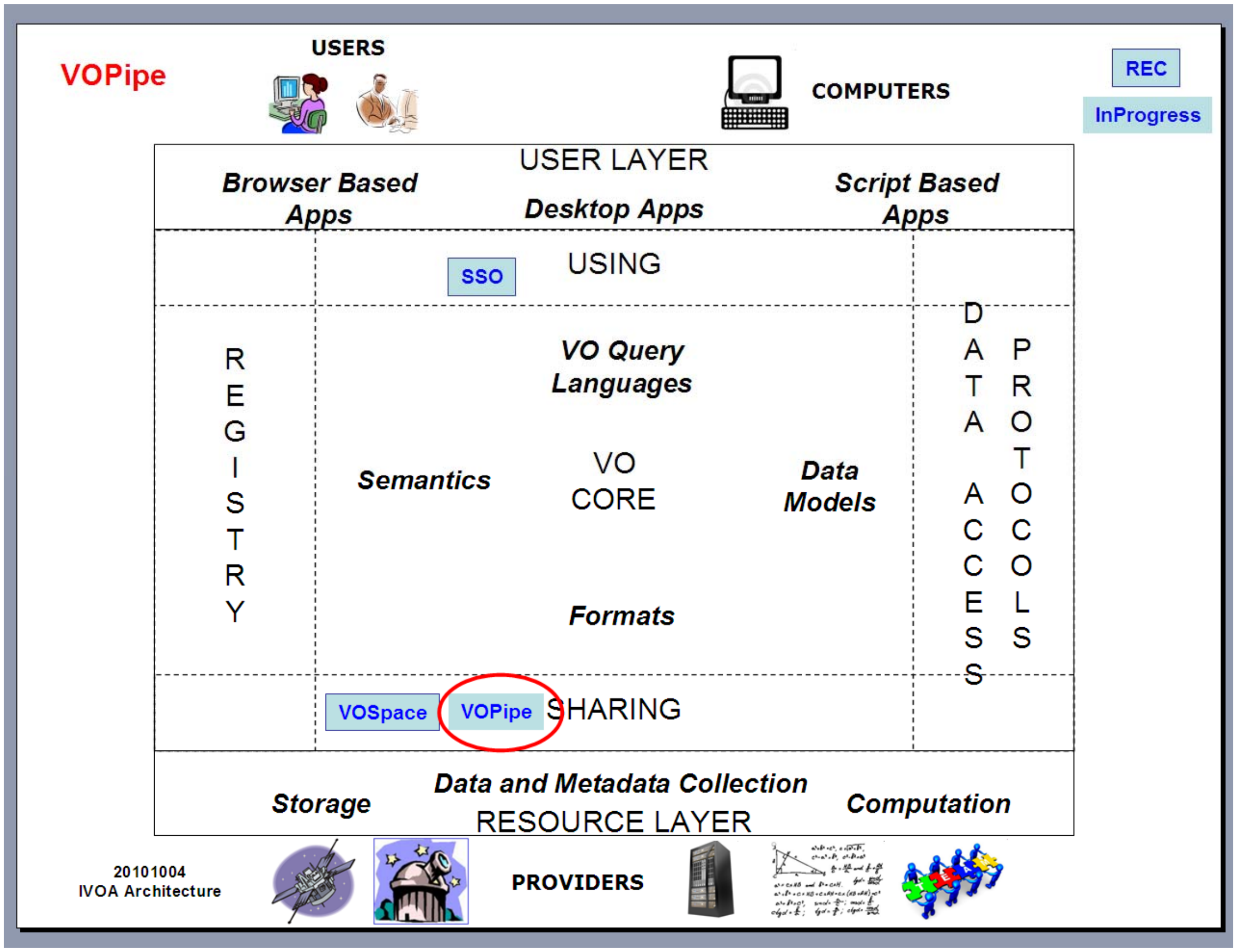

The VO provides a technical framework for the providers and the users to share their data and services ("Sharing"). VOSpace is the IVOA interface to distributed storage.

VOPipe builds on top of VOSpace and provides efficient data streaming between VOSpace instances. It is akin to reliable asynchronous messaging allowing multiple VOSpace instances to be chained together into data flows and chunks of data to be read from and written to VOSpace instances as required, e.g. facilitating data transport between VO query engines.

Where the access policy requires authentication of callers, the VOPipe service shall support the IVOA authentication methods such as Single Sign On profile (SSO).

In 2010, VOPipe is still under initial development.

## 11.3 UWS – Universal Worker Service

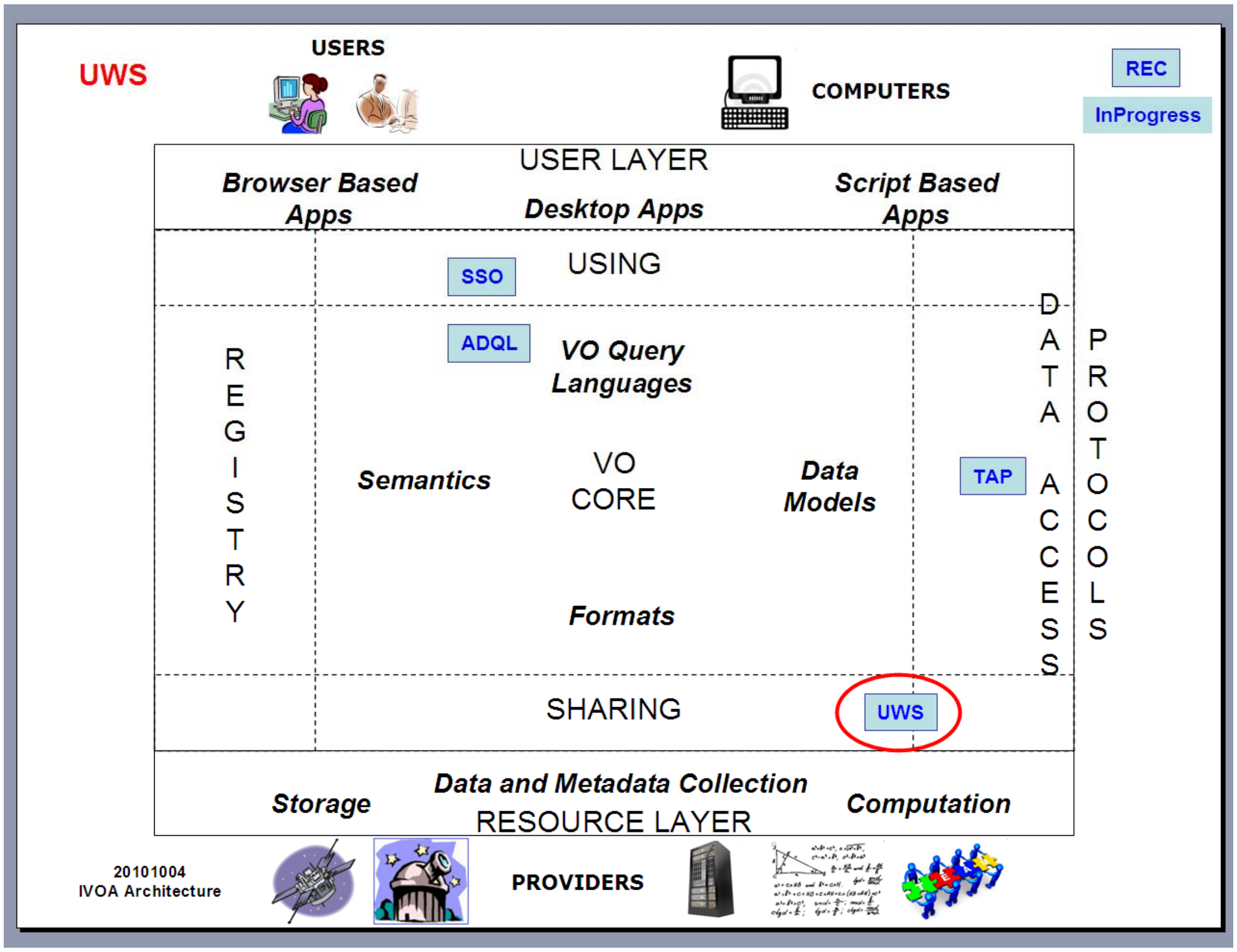

Simple VO web services are usually synchronous and stateless. Synchronous means that the client waits for each request to be fulfilled; if the client disconnects from the service then the activity is abandoned. Stateless means that the service does not remember results of a previous activity (or, at least, the client cannot ask the service about them). Synchronous, stateless services, in short, do not scale well.

Universal Worker Service (UWS) is a VO standard, being used by other standards and services to enable the development of VO applications managing asynchronous execution of jobs with state on VO services. UWS pattern allows a simple form of data sharing that is suitable for "workflow" situations and can be used by Data Access Services (currently TAP) or ADQL services. It utilizes IVOA standards for security (SSO) if it is desired that a non-public UWS be created.

UWS v1.0 became an IVOA Recommended standard in October 2010.